\documentclass{jpp}

\usepackage{float}
\usepackage{gensymb}
\usepackage{graphicx}
\usepackage{subcaption}
\usepackage[utf8]{inputenc}
\usepackage[T1]{fontenc}
\usepackage{amsmath}
\usepackage{hyperref}
\usepackage{mathrsfs}
\usepackage{color} %Only needed for the note commands added.

\hypersetup{
    colorlinks=true,
    linkcolor=blue,
    urlcolor=blue,
    citecolor=black
    }

\usepackage[toc,page]{appendix}
\newcommand{\figref}[1]{Figure~\ref{#1}}   
\newcommand{\V}[1]{\mathbf{#1}} 
\newcommand\Alfven{Alfv\'en }
\usepackage{color}

\shorttitle{Instabilities in Shocks}
\shortauthor{C. R. Brown, J. Juno, G. G. Howes, C. C. Haggerty, and S. Constantinou}

\title{Isolation and Phase-Space Energization Analysis of the Instabilities in Collisionless Shocks}

\author{C. R. Brown\aff{1}
  \corresp{\email{collin-brown@uiowa.edu}},
  J. Juno\aff{2},
  G. G. Howes\aff{1},
  C. C. Haggerty\aff{3},
  \and S. Constantinou\aff{3}}

\affiliation{\aff{1}Department of Physics and Astronomy, University of Iowa, Iowa City, IA 52240, USA
\aff{2}Princeton Plasma Physics Laboratory, Princeton, NJ 08543, USA
\aff{3}Institute for Astronomy, University of Hawai`i M\=anoa, Honolulu, HI 96822, USA}

\begin{document}

\maketitle

\begin{abstract}
We analyze the generation of kinetic instabilities and their effect on the energization of ions in non-relativistic, oblique collisionless shocks using a 3D-3V simulation by \texttt{dHybridR}, a hybrid particle-in-cell code. At sufficiently high Mach number, quasi-perpendicular and oblique shocks can experience rippling of the shock surface caused by kinetic instabilities arising from free energy in the ion velocity distribution due to the combination of the incoming ion beam and the population of ions reflected at the shock front. To understand the role of the ripple on particle energization, we devise the new instability isolation method to identify the unstable modes underlying the ripple and interpret the results in terms of the governing kinetic instability. We generate velocity-space signatures using the field-particle correlation technique to look at energy transfer in phase space from the isolated instability driving the shock ripple, providing a viewpoint on the different dynamics of distinct populations of ions in phase space. We generate velocity-space signatures of the energy transfer in phase space of the isolated instability driving the shock ripple using the field-particle correlation technique. Together, the field-particle correlation technique and our new instability isolation method provide a unique viewpoint on the different dynamics of distinct populations of ions in phase space and allow us to completely characterize the energetics of the collisionless shock under investigation.
\end{abstract}

\section{Introduction}

Collisionless shocks above a critical Mach number must invoke energy dissipation mechanisms other than resistivity to enable a steady-state structure in the transition region \citep{leroy1981simulation,wu1984microinstabilities}. Such supercritical shocks \citep{Edmiston:1984,Kennel:1985} can form when a large obstacle is put into a plasma flow that is traveling significantly faster than the fast magnetosonic velocity in the frame of the obstructing object. This disturbance creates a plasma wave that steepens until a shock forms, creating a transition from supersonic flow upstream to subsonic flow downstream. Collisionless shocks impact the dynamics and energetics of a plasma system in several ways, such as the ability to accelerate particles to high energy \citep{caprioli2018diffusive}, and the partitioning of energy between electrons and ions \citep{Savoini:2010}, which is allowed due to the lack of particle collisions to drive ion and electron temperatures to equilibrium.

For quasi-perpendicular and oblique shocks, a fraction of incoming ions are reflected at the shock transition and travel back upstream a distance that is typically of order one gyroradius, creating an unstable distribution of incoming and reflected ions that can generate unstable electromagnetic fluctuations which mediate the dissipation of energy and aid in forming the structure of the shock transition. This reflection plays a significant role in creating the foot/ramp/overshoot structure seen in the transverse (to the shock normal direction) magnetic field that is typical of collisionless shocks \citep{leroy1981simulation,leroy1982structure,Balogh:2013,Burgess:2015}. A gradual increase in the transverse magnetic field occurs in the foot, with a more rapid increase in the ramp of the shock. Beyond this, the overshoot increases the transverse component of the magnetic field above its eventual downstream asymptotic value.

For shocks with a Mach number and shock normal angle in a particular regime, the surface of the shock starts to ripple. This occurs as kinetic instabilities are driven by the unstable distribution created by the combination of the incoming and reflected populations of ions. This has been observed in simulation \citep{winske1988magnetic,mckean1995wave,lowe2003properties,burgess2016microstructure} and \textit{in-situ} with spacecraft \citep{johlander2016rippled,johlander2018shock}. Shock rippling has been studied previously using two-dimensional \citep{winske1988magnetic,mckean1995wave,lowe2003properties} and three-dimensional simulations \citep{burgess2016microstructure}. Studies that simulate only two spatial dimensions potentially risk suppressing some of the important degrees of freedom of the system \citep{burgess2007shock,zacharegkas2022kinetic}. Here we focus on the ion scale fluctuations caused by Alfv\'enic modes. The shock ripple may also impact other unstable fluctuations arising in the shock transition, for example the rippling may interact with whistler waves that propagate upstream and downstream \citep{gedalin_ganushkina_2022}.

Modeling this instability with the theory of linear waves in a homogeneous plasma has been difficult, as the instability that best matches the observations depends on the position relative to the shock transition. In the shock foot, the instability appears to be driven by streaming instabilities as the reflected population has a significant component of its velocity perpendicular to the velocity of the incoming stream, and this has been proposed to be due to either the modified two stream instability \citep{winske1988magnetic} or the ion Weibel instability \citep{burgess2016microstructure}. In the shock ramp, previous studies have concluded that the instability corresponds to the Alfv\'en ion cyclotron instability (AIC), as the waves generated locally tend to travel at the local Alfv\'en velocity and the plasma meets the temperature anisotropy threshold that are also required by the AIC instability \citep{Gary:1976,winske1988magnetic,lowe2003properties,klein2015predicted}. Agreement between linear theory with the AIC instability has been found \citep{mckean1995wave}, but the analysis methods used make it difficult to isolate instabilities in a restricted time or space domain. This difficulty is further enhanced in higher energy shocks in which the instabilities excited can cause non-stationary, non-steady state behavior such as shock reformation because of how rapidly these instabilities evolve in space and time. Furthermore, modeling this instability is also complicated by the non-homogeneous nature of the plasma through the shock transition.

Previous work has compared the linear theory of kinetic instabilities in homogeneous plasmas with two-dimensional shock simulations \citep{winske1988magnetic,mckean1995wave,burgess2007shock}. These studies tracked the shock as a function of time to measure the growth of any expected instabilities, or Fourier transformed a block of the simulation in space and time, which loses the locality of the dynamics of the shock along the shock normal direction and the locality of the plasma parameters. While these attempts have been successful at examining the dynamics in the ramp, to our knowledge, no direct comparisons between linear theory and the instabilities in the foot have been made. Only quantitative similarities between properties of both the instabilities in the foot and linear instabilities that are suspected to be present have been found \citep{burgess2016microstructure}. This motivates the need for development of a more versatile method for comparing simulations to linear theory, as such a tool will have applications beyond analysis of the instabilities causing shock ripple.

The purpose of this paper is to develop tools for analyzing how kinetic instabilities affect the dynamics and energetics of a collisionless shock and to demonstrate these tools on the main instability causing the surface of the shock to ripple. We develop a method that can locally identify the kinetic instability (or instabilities) present and characterize its properties from simulation data by analyzing the `fluctuating fields', as defined in Section \ref{sec:id}. We show that these fluctuations are consistent with the properties of wave modes from the linear Vlasov-Maxwell dispersion relation based on the local plasma parameters at that position, as shown in \figref{fig:kawsweeps}. Furthermore, we use this definition of the fluctuating fields to compute the energization due to the fluctuations arising from kinetic instabilities in the shock transition. Thus, we devise a means to separate the approximately steady-state bulk shock energization of particles due to the shock transition from the energization due to the instability-driven fluctuations, with the aim to quantify and explain how the electromagnetic fluctuations arising from kinetic instabilities affect the energization of particles through the shock transition.

In this paper, we will present the novel instability isolation method and apply it to a simulation of an oblique collisionless shock. We use this method in conjunction with the field-particle correlation technique to produce the velocity-space signatures of particle energization in phase space due to an isolated instability. We use these methods to investigate the kinetic instability responsible for the shock ripple and assess the impact of the instability on the ion energization in the shock. These methods allow for one to separate the energization of particles due to the steady-state physics of the shock from that due to the kinetic instabilities arising in the shock transition. In Section~\ref{sec:simsetup}, we describe the setup and details of the 3D-3V \texttt{dHybridR} simulation of a oblique collisionless shock. Section~\ref{sec:id} presents the instability isolation method devised to analyze the properties of the modes driven unstable by kinetic instabilities in the shock transition. We analyze the particle energization in the ramp of this simulated shock, separating the energization due to the bulk steady-state shock physics (a transverse-plane average of the fields through the shock) from the energization due to the kinetic instabilities responsible for the shock ripple in Section 4. We conclude in Section~\ref{sec:conc} by discussing new avenues of investigation made possible by these techniques to understand the dynamics and energetics of other non-stationary shocks.

\section{Simulation Setup}
\label{sec:simsetup}
We present fully three-dimensional hybrid simulations using the massively parallel code \texttt{dHybridR} \citep{gargate2007dhybrid,haggerty2019dhybridr}. We produce a shock by sending a supersonic flow in the  $-\hat{x}$ direction towards a reflecting wall at $x=0$. The resulting colliding flows generate a shock that propagates in the $+\hat{x}$ direction in the simulation domain. We maintain periodic boundary conditions in the $\hat{y}$ and $\hat{z}$ directions. We use $1000$ particles per cell in the initial state of each cell and at the upstream boundary where particles are continuously injected. We normalize the magnetic fields and number density to their values in the upstream region, $B_0$ and $n_0$. Lengths are scaled to the upstream ion inertial length $d_{i,0} \equiv c / \omega_{pi,0}$, where $\omega_{pi,0} = \sqrt{4 \pi n_0 e^2/m_i}$ is the ion plasma frequency using the upstream density. Time is scaled to the inverse ion gyrofrequency $\Omega_{i,0}^{-1} \equiv c m_i / e B_0$ based on the upstream magnetic field. Velocity is normalized using to the upstream Alfv\'en velocity, $v_{A,0} = B_0/\sqrt{4 \pi n_0 m_i} = d_{i,0} \Omega_{i,0}$. Electric fields are normalized to $B_0 v_{A,0} / c$. The ratio of the \Alfven velocity to the speed of light is $v_{A,0}/c = 1/125$. The plane normal to the shock propagation direction, the transverse plane, has an area of $L_y \times L_z = 12 \, d_{i,0} \times 12 \, d_{i,0}$. The length of the simulation along the direction of the shock normal is $L_x = 98.75 \, d_{i,0}$. The simulation is divided into square cells with sides of length $0.25 \, d_{i,0}$. We set the initial state of the inflowing plasma to have beta values $\beta_i = \beta_e = 1$, where the species $s$ plasma beta is given by $\beta_s = 8 \pi n_s T_{s,0}/B_0^2$ and temperatures are expressed in units of energy (absorbing the Boltzmann constant into $T_{s,0}$). The upstream ion thermal velocity is defined by $v_{ti,0}= \sqrt{2 T_{i,0}/m_i}$. We inject particles at the upstream boundary with velocity $U_{inj} = -6 \, v_{A,0} \, \hat{x}$ in the frame of the simulation box and impose an initial magnetic field with $\theta_{B_n} = 45 \degree$ with $\mathbf{B}_0 = (1/\sqrt{2})B_0 \, \hat{x} + (1/\sqrt{2})B_0 \, \hat{z}$.

\begin{figure} 
    \centering
    \includegraphics[width=.85\textwidth]{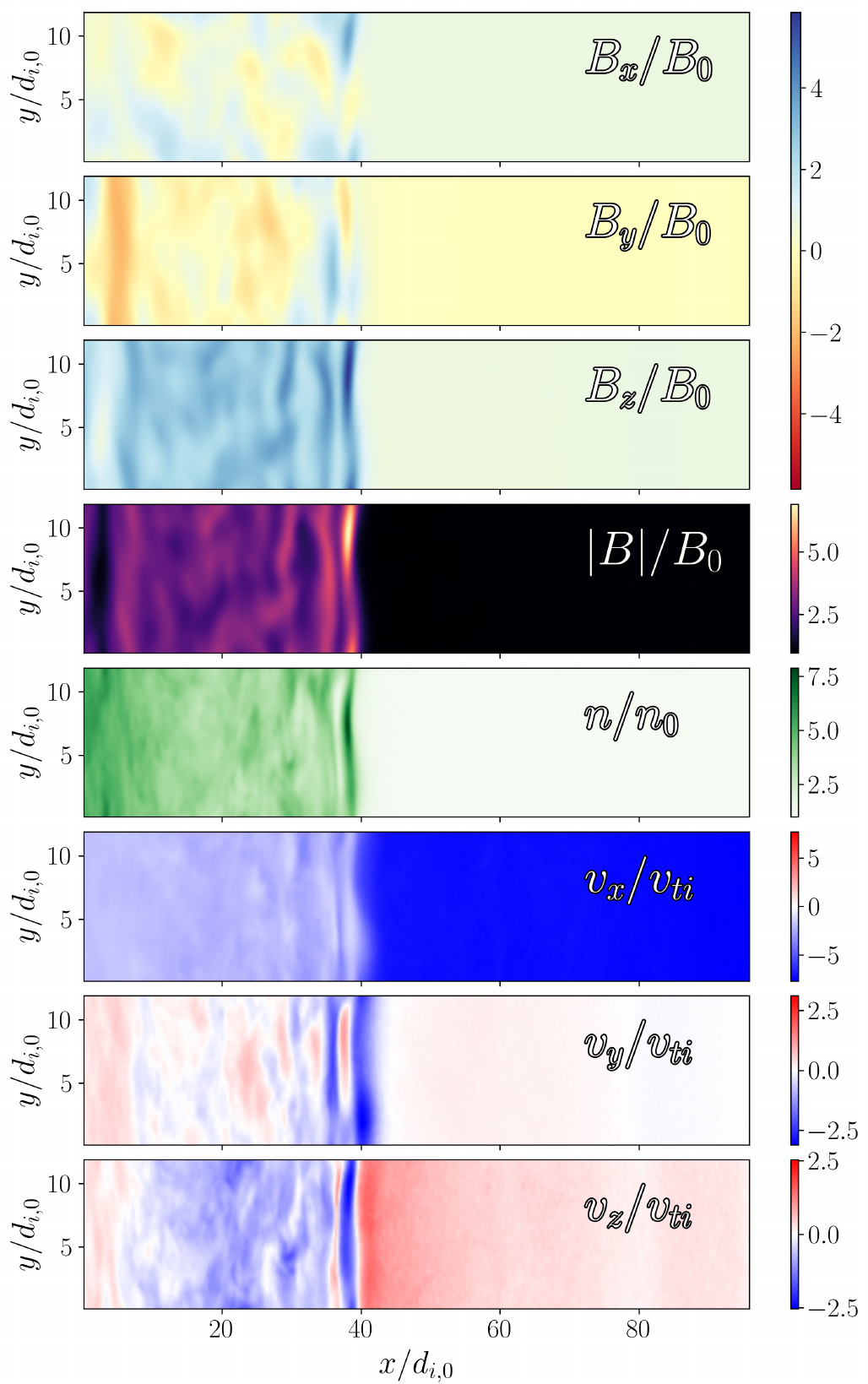}
    \caption{2D slice of magnetic fields and fluid moments of a \texttt{dHybridR} simulation with shock velocity of $M_A=7.88$ (in the downstream rest frame) and $\theta_{B_n} = 45 \degree$ shock at $\Omega_{i,0} t = 20$. At this time, the shock is at $x / d_{i,0} \simeq 40$.}
    \label{fig:pmeshsuperplot}
\end{figure}

\begin{figure} 
    \centering
    \includegraphics[width=.87269\textwidth]{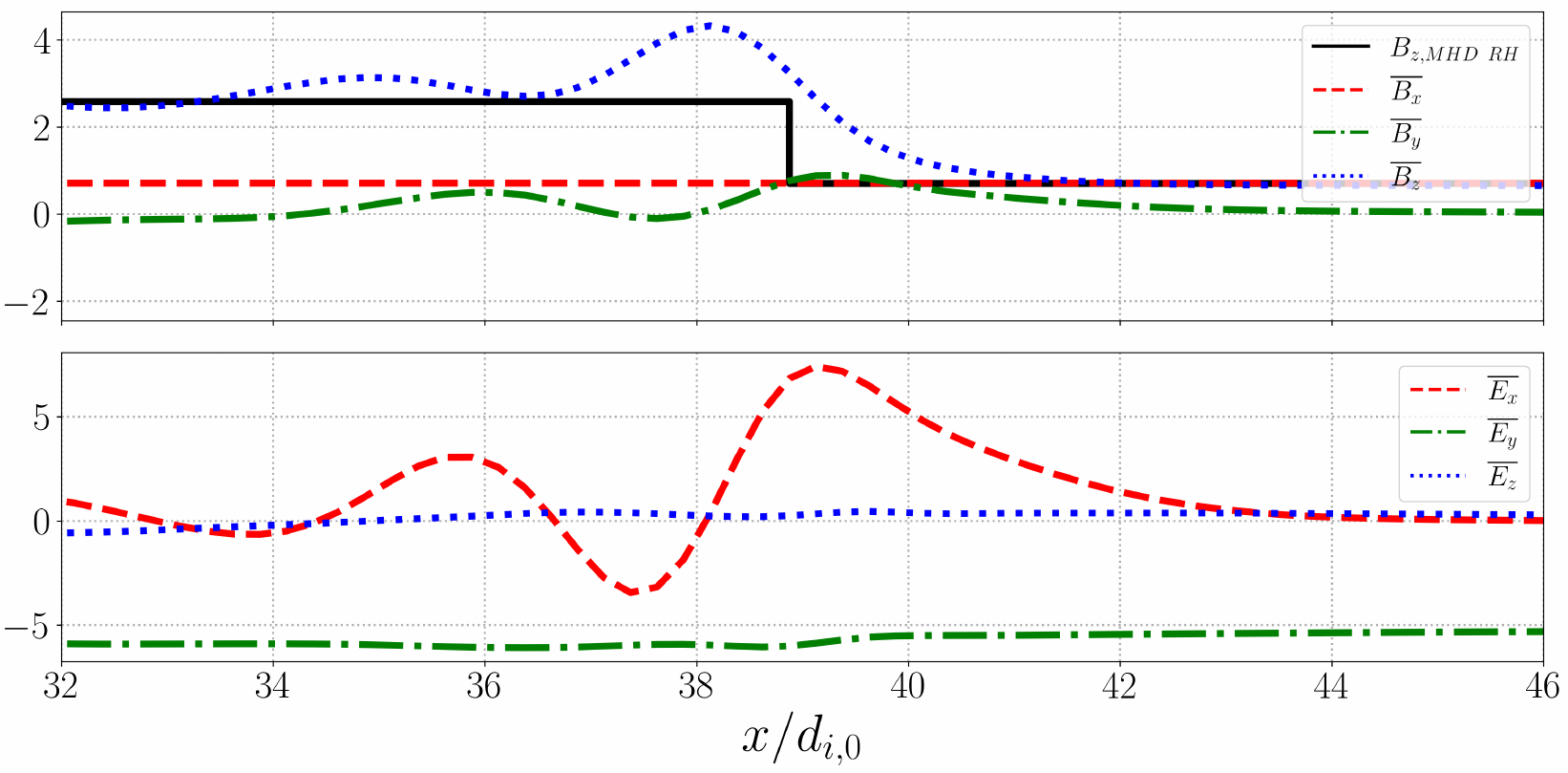}
    \caption{Average (over the transverse plane) magnetic fields (\emph{top}) and electric fields (\emph{bottom}) from the hybrid simulation at $\Omega_{i,0} t = 20$.  The black curve shows the transverse magnetic field jump $B_{z,2}/B_{z,1}=3.66$ predicted by the MHD Rankine-Hugoniot jump conditions for  a collisionless shock with $M_A=7.88$, $\theta_{B_n} = 45^\circ$, and $\beta=2$. }
    \label{fig:fieldscloseup}
\end{figure}

\begin{figure} 
    \centering
    \includegraphics[width=1.\textwidth]{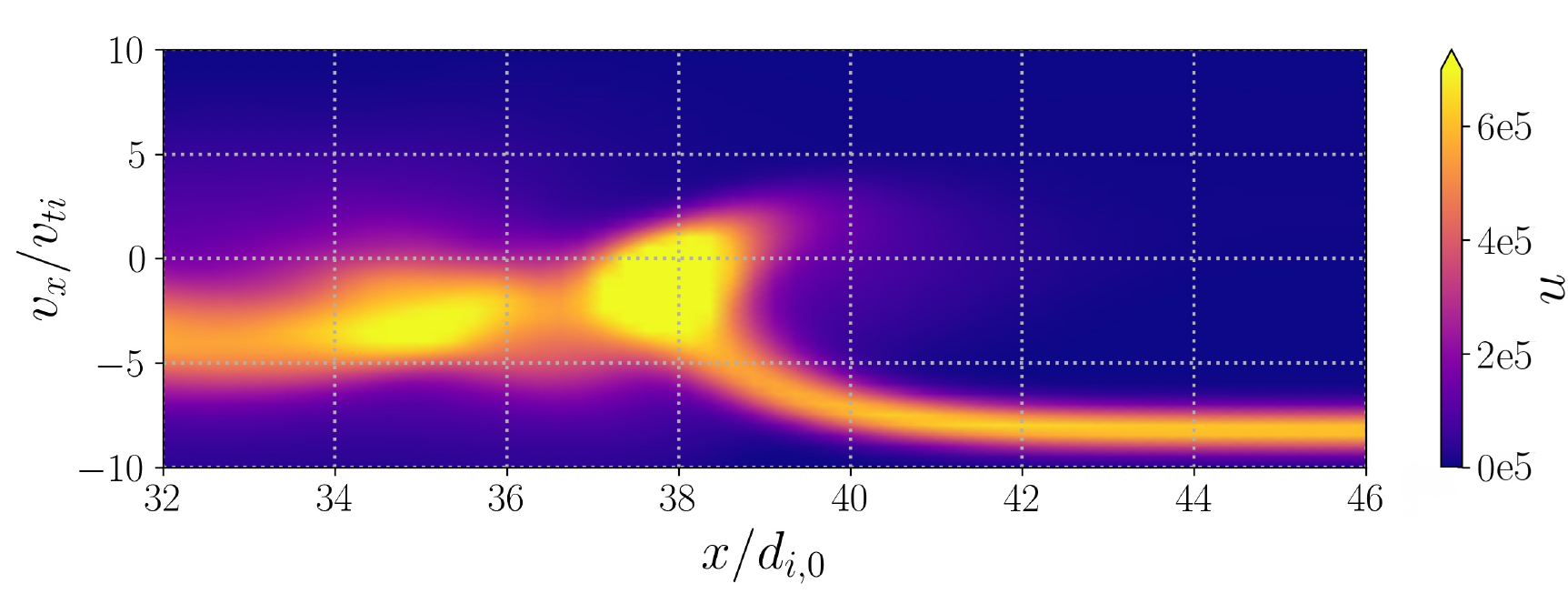}
    \caption{Ion distribution function $f_i(x,v_x)$, integrated over $v_y$ and $v_z$, in the shock-rest frame. Ions are reflected back upstream a distance of order one $d_{i,0}$.}
    \label{fig:histxxvy}
\end{figure}

Figure \ref{fig:pmeshsuperplot} shows the magnetic field structure, ion density, and bulk flow velocity at $\Omega_{i,0} \, t = 20$. The incoming flow in the shock-rest frame (as inferred from the simulation) has an Alfv\'en Mach number $M_A=7.88$, and the shock does not experience any shock reformation or significant `breathing', i.e. the shock velocity is stable throughout the simulation. We track the shock in the simulation frame as discussed in appendix \ref{appendix:shocktrack}, and Lorentz transform the electromagnetic fields and ion velocity distributions to the shock-rest frame. The electric and magnetic fields averaged across the transverse plane (in the shock-rest frame) are shown in Figure \ref{fig:fieldscloseup}. At $\Omega_{i,0} t = 20$, the shock has reached a position $x/d_{i,0} \simeq 40$. The compressible component of the magnetic field, $B_z$, has a compression ratio of $r \approx 3.5$. This ratio is in approximate agreement with the predicted compression ratio of $r = 3.62$ using MHD Rankine-Hugoniot jump conditions (see Appendix~\ref{appendix:shocktrack}). In this parameter regime and at this time in the simulation, MHD and kinetic shocks are in good agreement, but simulations of kinetic shocks can deviate \citep{bret2020can,haggerty2020kinetic,caprioli2020kinetic}. We observe ions being reflected in the phase-space plot $f_i(x,v_x)$, shown in Figure \ref{fig:histxxvy}, creating the instability discussed in Section \ref{sec:id}. The distribution of ions evolves rapidly along the $\hat{x}$ direction near the shock transition region, creating a small range in $x$ where the distribution is unstable, as discussed in Section~\ref{sec:ceivsx}.

\begin{figure} 
    \centering
    \includegraphics[width=\textwidth]{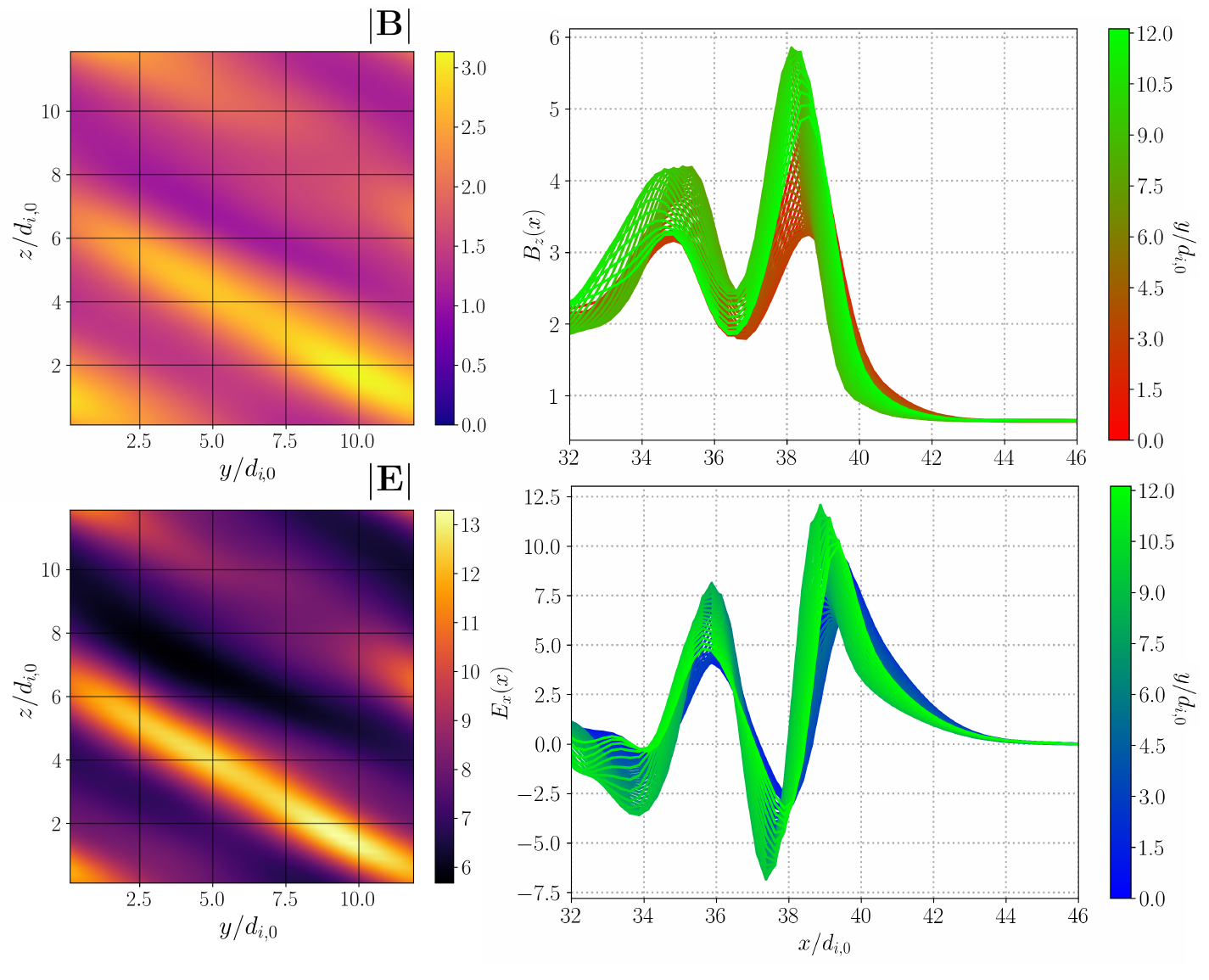}
    \caption{2D slice of the total magnetic field, $|\mathbf{B}(x_0,y,z)|$, at $x_0/ d_{i,0} = 39.875$ (\emph{Top Left}) and the total electric field, $|\mathbf{E}(x_0,y,z)|$ (\emph{Top Right}) over the transverse plane. There is rippling of the shock with $k_y \, d_{i,0} = -0.52$ and $k_z \, d_{i,0} = -1.05$. Line slices of the compressible magnetic field component, $B_z(x,y_i,z_0)$ (\emph{Bottom Left}) and $E_x(x,y_i,z_0)$ (\emph{Bottom Right}) as a function of $x$ with a fixed value of $z_0/d_{i,0} = 0.125$ for a set of discrete values of $y_i/d_{i,0}$. }
    \label{fig:2dbzpmesh}
\end{figure}

There is rippling in the shock at $x / d_{i,0} = 39.875 $. This location corresponds to the most upstream extent of the ramp. Figure \ref{fig:2dbzpmesh} shows a 2D slice of the rippling over the transverse plane. We see a periodic rippling in the magnetic field, indicating the presence of an instability launching a wave that is rippling the location of the shock front across the transverse direction. The width of the ripple is estimated to be $\Delta x /d_{i,0} \simeq 1$ from Figure \ref{fig:2dbzpmesh}.  There is no variation in the transverse plane in the upstream region, showing that the unstable ion velocity distribution arising in the shock transition induces instabilities which produce a variation of the fields in transverse directions.

\section{Instability Isolation Method}
\label{sec:id}
To isolate the unstable fluctuations from the variation of the electromagnetic fields due to the steady-state physics of the shock transition in our collisionless shock simulations, we first separate the average of these fields over the transverse plane of the shock (\emph{i.e},. along the shock face) from the fluctuations across that plane, where the boundary conditions in the plane are periodic.  We define the \emph{transverse-plane averaged fields} (hereafter denoted the \emph{averaged fields} for brevity) by
\begin{equation}
\overline{\mathbf{E}}(x) \equiv \frac{1}{L_y L_z}\int_0^{L_y}dy \int_0^{L_z}dz  \ \mathbf{E}(x,y,z).
\label{eq:avgfields}
\end{equation}
The \emph{fluctuating fields} are then computed by
\begin{equation}
\delta \mathbf{E}(x,y,z) = \mathbf{E}(x,y,z)-\overline{\mathbf{E}}(x),
\label{eq:fluctfields}
\end{equation}
where $\mathbf{E}(x,y,z)$ is the total electric field from the simulation in the shock-rest frame.  The motivation for this separation of the average and fluctuating fields is that the variation of the fields and flows associated with compression through the shock transition, in the absence of instabilities or other time variation in the shock-rest frame, is inherently one-dimensional along the shock normal\footnote{Note that, in principle, an instability could lead to an unstable mode with a wavevector strictly along the normal direction, so the variation due to that mode would be included in our determination of the averaged fields; however, if one or both of the transverse directions are included in a shock simulation, it seems extremely unlikely that an instability would arise with variation solely along the normal direction, so our separation of averaged and fluctuating fields is robust as long as there is some nonzero transverse variation of the unstable modes.}.
We denote this one-dimensional physics of the averaged fields the \emph{steady-state shock physics}, and the dynamics and energetics of the unstable modes is the \emph{instability physics}. Note that the one-dimensional steady-state shock physics is only approximately steady-state---for supercritical shocks, the structure through the shock transition does undergo slight oscillations in time in the shock-rest frame (sometimes denoted ``breathing'' of the shock), oscillations which increase in amplitude as the Mach number increases. Upon reaching or exceeding the second critical Mach number, $M_{2c,A} \approx 15.1$ \citep{krasnoselskikh2002nonstationarity, oka2006whistler}, these oscillations lead to the phenomenon of shock reformation, but as this simulation has a significantly lower \Alfven Mach number than this second critical threshold, $ M_A< M_{2c,A}$, we expect no reformation in this simulation.

It should be noted that despite our separation of the steady-state shock physics from the instability physics using equations \eqref{eq:avgfields} and \eqref{eq:fluctfields}, these phenomena are causally related.
That is, the steady-state shock physics, which has no inherent variation in the transverse plane, generates the unstable ion velocity distributions that give rise to the instabilities, so the energetics of the steady-state shock physics is not decoupled from the instability physics, but rather drives those instabilities.

The basic steps of the instability isolation method, using the electromagnetic fields in the shock-rest frame, are:
\begin{enumerate}
    \item Compute the averaged electric field $\overline{\mathbf{E}}(x)$ using \eqref{eq:avgfields} and the fluctuating electric field $\delta \mathbf{E}(x,y,z)$ using \eqref{eq:fluctfields}.
    \item Fourier transform the fluctuating fields over the transverse plane $(y,z)$ to obtain $\widetilde{\delta \mathbf{E}} (x,k_y,k_z)$.
    \item Perform a wavelet transform along the shock normal direction to obtain the Wavelet-Fourier Transform (WFT) $\delta \hat{\mathbf{E}} (k_x,k_y,k_z;x)$.
    \item Use the same procedure as in (i)--(iii) above on the magnetic field to obtain the WFT of the magnetic field, $\delta \hat{\mathbf{B}} (k_x,k_y,k_z;x)$.
    \item At the position $x=x_0$ along the normal direction, compute the local averaged total magnetic field, $\overline{\mathbf{B}}(x_0)$. 
    \item Generate a local magnetic field-aligned coordinate (FAC) system
    $(\hat{\mathbf{e}}_{\perp 1},\hat{\mathbf{e}}_{\perp 1},\hat{\mathbf{e}}_{\parallel})$ using the normal direction $\hat{\mathbf{n}}=\hat{x}$ and $\overline{\mathbf{B}}(x_0)$.
    \item Rotate the fluctuating WFT fields $\delta \hat{\mathbf{E}} (k_x,k_y,k_z;x_0)$ and $\delta \hat{\mathbf{B}} (k_x,k_y,k_z;x_0)$ into the FAC system.
    \item To estimate the frequency of an unstable (local) plane-wave mode given by $(k_{\perp 1},k_{\perp 2},k_\parallel;x_0)$ from the WFT transform, use Faraday's Law in the FAC coordinate system.
    \item Finally,  compare the estimated frequency of the unstable mode from Faraday's Law to the frequencies of the different wave modes from a linear dispersion relation solver.
\end{enumerate}

%----------------------------------------------------------------
\subsection{Wavelet-Fourier Transform}
Since instabilities typically are dominated by one or a few of the most rapidly growing unstable modes, our first task is to transform the fluctuating fields into local plane-wave modes using a combined Wavelet-Fourier transform (WFT).  Since the boundary conditions in the transverse plane $(y,z)$ are periodic, the fluctuating electric field $\delta \mathbf{E}(x,y,z)$ and fluctuating magnetic field $\delta \mathbf{B}(x,y,z)$  are Fourier transformed over these two directions to obtain the complex Fourier coefficients as a function of $(x,k_y,k_z)$, yielding the transformed fields
$\widetilde{\delta \mathbf{E}} (x,k_y,k_z)$ and $\widetilde{\delta \mathbf{B}} (x,k_y,k_z)$. 

Next, since the steady-state shock physics leads to significant non-periodic variations in the normal direction, we employ the wavelet transform \citep{torrence1998practical} along the shock normal, $\hat{\mathbf{n}}=\hat{x}$, 
\begin{equation}
\delta \hat{\mathbf{E}}(k_x,k_y,k_z;x) \equiv W_\psi \{ \widetilde{\delta \mathbf{E}}(x,k_y,k_z)\} = \sqrt{|k_x|} \int_{-\infty}^{\infty} dx^\prime \, \, \psi^*_{\sigma_0}\big[(x^\prime-x) k_x\big]  \widetilde{ \delta\mathbf{E}}(x',k_y,k_z)
\end{equation}
with the complex Morlet wavelet
\begin{equation}
    \psi_{\sigma_0}(x)=e^{i k_x x} e^{-\frac{1}{2}\big(\frac{x k_x}{\sigma_0}\big)^2} \pi^{\frac{1}{4}} \sqrt{\frac{k_x}{\sigma_0}}
\label{eq:Morletwavelet}
\end{equation}
where $\sigma_0$ is a parameter that allows a trade off between resolution in position and wavenumber \citep{najmi1997continuous} and $\widetilde{\delta \mathbf{E}}(x,k_y,k_z)$ is the Fourier transformed fluctuating electric field. At higher $\sigma_0$, wavenumber is better constrained at the cost of spatial resolution. It should be noted that we normalize the result at each wavenumber in order to maintain the relative energy of each mode, as suggested in \cite{torrence1998practical}. By combining the Fourier transform of the perturbed fields in the transverse plane with the wavelet transform in the normal direction, we are able to determine the approximate wavevector of unstable fluctuations locally at position $x$, enabling us to explore the properties of the unstable modes as they vary in the normal direction through the shock.

\begin{figure} 
    \centering
    \includegraphics[width=\textwidth]{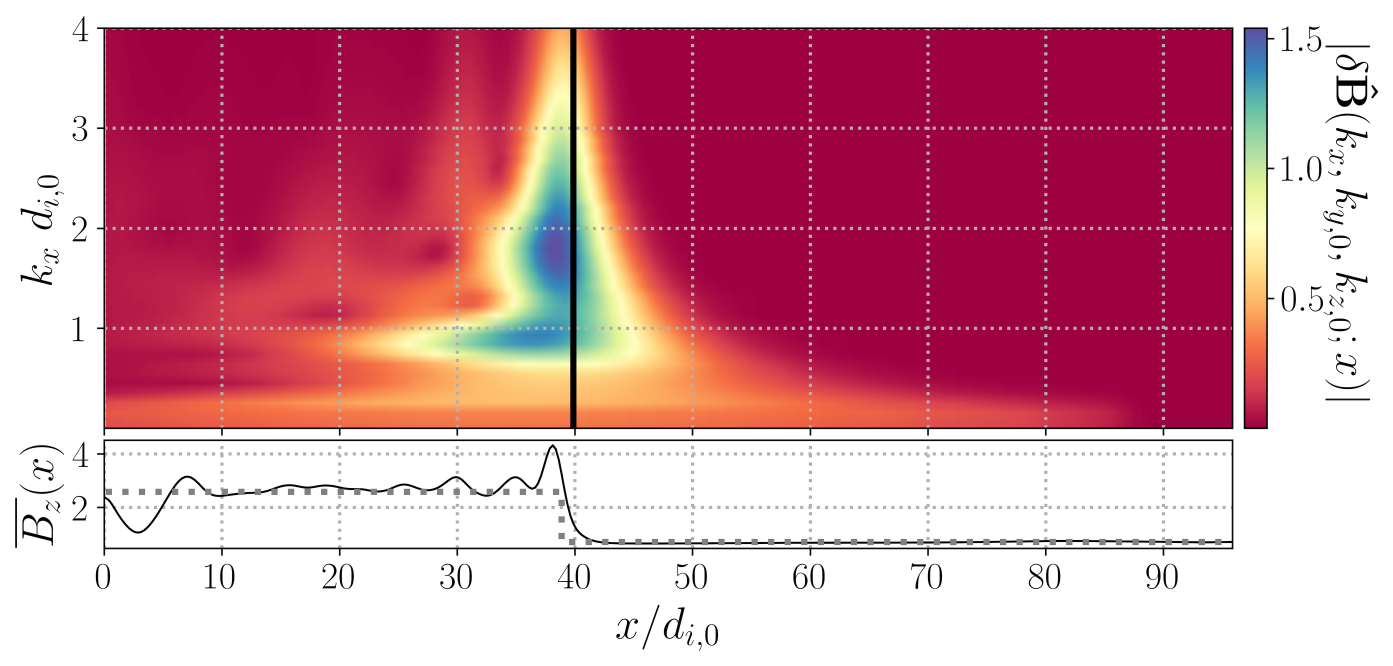}
    \caption{(\emph{Top:}) Wavelet transform of the magnetic field fluctuations, $\delta \hat{\mathbf{B}}$, of an oblique shock with a shock velocity of $M_A \approx 7.88$ and a shock normal of $\theta_{B_n} = 45 \degree$ with fixed $k_{y,0} = -0.52$ and $k_{z,0} = -1.05$ (i.e. $\delta \hat{\mathbf{B}}(x;k_x,k_{y,0},k_{z,0})$) at time $t=20 \, \, \Omega_{i,0}^{-1}$. A vertical black line indicates the position $x/d_{i,0}=39.875$ at which we determine the dominant wavemode. We see larger values for $| \delta \hat{\mathbf{B}}|$ in the ramp and overshoot of the shock. (\emph{Bottom:}) Compressible magnetic field component, $\overline{B}_z$, for reference.}
    \label{fig:wlt1}
\end{figure}

The same WFT is applied to obtain the local wavevectors of the perturbed magnetic field as a function of the position along the shock normal, $\delta \hat{\mathbf{B}}(k_x,k_y,k_z;x)$.
Figure \ref{fig:wlt1} shows that the unstable mode amplitudes peak in the shock transition region, rather than in the upstream or downstream regions. In this figure, we plot the wavelet transform of the dominant transverse Fourier mode $(k_y d_{i,0},k_z d_{i,0})=(-0.52,-1.05)$ at position $x_0/d_{i,0}=39.875$, as shown in Fig.~\ref{fig:knormBpmesh}(c). $|\delta \hat{\mathbf{B}}(x_0=39.875d_{i,0})|$ assumes a maximum at $(k_x d_{i,0},k_y d_{i,0},k_z d_{i,0})= (1.62,-0.52,-1.05)$. There exists nonzero amplitude of $|\delta \hat{\mathbf{B}}(x_0=39.875d_{i,0})|$ far upstream at low $k_x$, although this may be due to the inability of the wavelet transform to localize waves with small wavenumber.
Most critically, we note that the dominant transverse Fourier mode $(k_y d_{i,0},k_z d_{i,0})=(-0.52,-1.05)$ matches the observed ripple structure seen in Figure~\ref{fig:2dbzpmesh}. We will further motivate our selection of this wave mode in Section~\ref{sec:FAC}, explain how to measure frequency using WFT in Section~\ref{sec:freqest}, and compare our measured wave to dispersion relations from gyrotropic linear theory in Section~\ref{sec:comp2lin} to show that we are able to isolate and identify the wave present in this simulation.

%----------------------------------------------------------------
\subsection{Magnetic Field-Aligned Coordinate (FAC) System}
\label{sec:FAC}

We construct a local magnetic field aligned coordinate (FAC) system at $x_0/d_{i,0} =39.875$, defined by the orthonormal basis
$(\hat{e}_{\perp 1},\hat{e}_{\perp 2},\hat{e}_{\parallel})$, where
\begin{equation}
\hat{e}_{\parallel} = \frac{\overline{\mathbf{B}}(x_0)}{|\overline{\mathbf{B}}(x_0)|}
\end{equation}
\begin{equation}
\hat{e}_{\perp 2} = 
\frac{\overline{\mathbf{B}}(x_0) \times \hat{\mathbf{n}}}
{|\overline{\mathbf{B}}(x_0) \times \hat{\mathbf{n}}|}
=\frac{\overline{\mathbf{B}}(x_0) \times \hat{x}}
{|\overline{\mathbf{B}}(x_0) \times \hat{x}|}
\end{equation}
\begin{equation}
\hat{e}_{\perp 1} = 
\frac{\hat{e}_{\perp 2} \times \hat{e}_{\parallel}}
{|\hat{e}_{\perp 2} \times \hat{e}_{\parallel}|}.
\end{equation}
At $x_0$, we measure the averaged magnetic field $\overline{\mathbf{B}}(x_0)/B_0 = (0.707, 0.698, 1.390)$, which notably has a substantial nonzero component in the $\hat{y}$ direction.

\begin{figure} 
    \centering
    \includegraphics[width=1.\textwidth]{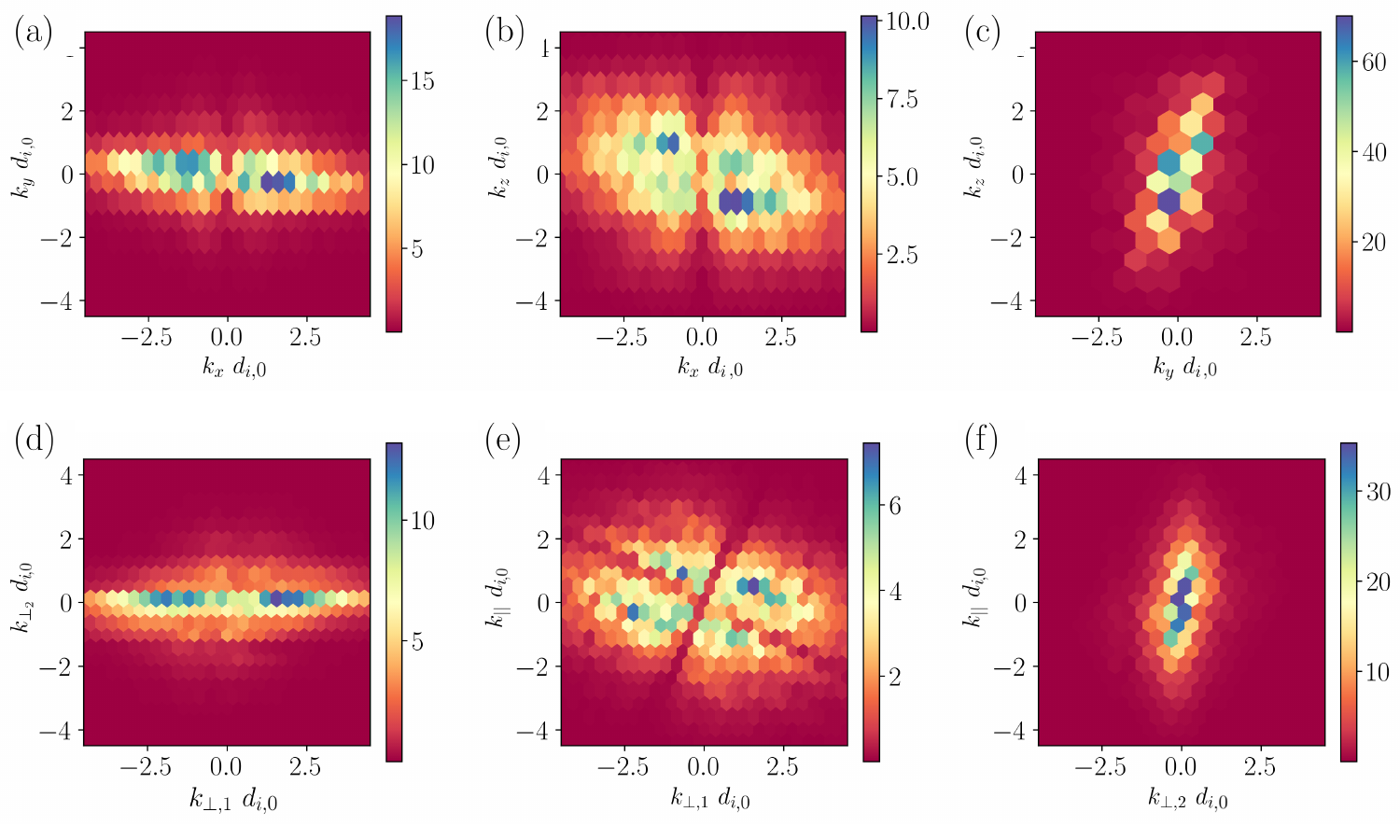}
    \caption{Projections of $|\delta \hat{\mathbf{B}}(k_{x},k_{y},k_{z})|$ (\emph{Top Row}) and  $|\delta \hat{\mathbf{B}}(k_{||},k_{\perp 1},k_{\perp 2})|$ (\emph{Bottom Row}) in the shock ramp at $x_0/d_{i,0}=39.875$ using hexagonal binning due to the non-uniformity of data in field aligned coordinates. There is a dominant wave mode in the FAC system with $(k_{\perp 1}d_{i,0},k_{\perp 2}d_{i,0},k_{||}d_{i,0}) = (1.97,0.00,0.34)$, corresponding to $(k_{x}d_{i,0},k_{y}d_{i,0},k_{z}d_{i,0}) = (1.62,-0.52,-1.05)$ in the simulation coordinates, along with its conjugate mode due to the reality condition. There is little power off the $k_{\perp 2}$ axis. The `plus' structure in the middle panel, bottom row, arises due to our projection of a grid in simulation coordinates $(x,y,z)$ onto our FAC system.}
    \label{fig:knormBpmesh}
\end{figure}

Figure \ref{fig:knormBpmesh} shows a singular dominant WFT wave mode (with conjugate pair) around $(k_x d_{i,0},k_y d_{i,0},k_z d_{i,0}) = (1.61,-0.52,-1.05)$, which corresponds to $(k_{\perp 1}d_{i,0},k_{\perp 2}d_{i,0}, k_{||}d_{i,0}) = (1.97,0.00,0.34)$ in our FAC system. The localization of amplitude around $k_{\perp 2}=0$ suggests that our FAC system is a natural coordinate system for describing the unstable mode underlying the shock ripple. Note that the transverse Fourier wavevector of this dominant unstable mode $(k_y d_{i,0},k_z d_{i,0})=(-0.52,-1.05)$ is consistent with the  dominant plane wave structure seen in the total magnetic field seen in Figure \ref{fig:2dbzpmesh}. 

%----------------------------------------------------------------
\subsection{Frequency Estimation Using Faraday's Law}
\label{sec:freqest}

Since the dominant wave mode in the FAC system has $k_{\perp 2}=0$, we take the wavevector to lie in the $(\hat{e}_{\perp 1},\hat{e}_{\parallel})$ plane, such that $\mathbf{k} = k_{\perp 1} \hat{e}_{\perp 1} + k_\parallel \hat{e}_{\parallel}$.  Assuming local plane-wave modes that vary as $\exp(i \mathbf{k} \cdot \mathbf{x} - i \omega t)$, we Fourier transform Faraday's law in time and space to relate $\omega \mathbf{B} = \mathbf{k}\times \mathbf{E}$.  We convert these equations into a dimensionless normalization for the complex WFT coefficients using $\mathbf{B}' = \delta \hat{\mathbf{B}}/B_0$, $\omega' = \omega/\Omega_i$, and $\mathbf{E}' = \delta \hat{\mathbf{E}}/(v_A B_0)$, where $B_0$ is the magnitude of the upstream magnetic field and we use a dimensionless wavevector $\mathbf{k} \, d_i$, noting that $d_i = v_A/\Omega_i$ should be the \emph{local} value of the ion inertial length when comparing to linear wave modes. Thus, the components of Faraday's Law may be expressed in this dimensionless normalization by
\begin{equation}
  \omega'  B'_{\perp 1} = -k_\parallel d_i  E'_{\perp 2}
\label{eq:faraday1}
\end{equation}
\begin{equation}
  \omega' B'_{\perp 2} =k_\parallel d_i E'_{\perp 1}
\label{eq:faraday2} - k_{\perp 1} d_i E'_\parallel 
\end{equation}
\begin{equation}
  \omega' B'_{\parallel} = k_{\perp 1} d_i E'_{\perp 2}
\label{eq:faraday_par}
\end{equation}

The complex WFT coefficients for the dimensionless fluctuating
electric field $\mathbf{E}'$ and magnetic field $\mathbf{B}'$ from the
simulation for the dominant wave mode are presented in table \ref{tab:wvmdtable}. We can use
equations (\ref{eq:faraday1}), (\ref{eq:faraday2}), and
(\ref{eq:faraday_par}) to obtain three separate determinations of the
real frequency $\omega$, also presented in table \ref{tab:wvmdtable}. Note that, since the original fields $\mathbf{E}$ and $\mathbf{B}$ before the WFT transform are real, the complex WFT coefficients must satisfy the reality condition $\hat{\V{E}}(\V{k})=\hat{\V{E}}^*(-\V{k})$ and $\hat{\V{B}}(\V{k})=\hat{\V{B}}^*(-\V{k})$, so combining the coefficients for a given $\V{k}$ and its conjugate $-\V{k}$ yields strictly real fields, and therefore we obtain a real value for $\omega$ from each equation.  

\begin{table}
 \begin{center} \begin{tabular}{|l||c|c|c||c|c|c||c|c|c|}
    \hline
    & $E'_{\perp 1}$&$E'_{\perp 2}$& $E'_\parallel$& $B'_{\perp 1}$&$B'_{\perp 2}$&$B'_{\parallel} $&$\omega$ (\ref{eq:faraday1})&$\omega$ (\ref{eq:faraday2})&$\omega$ (\ref{eq:faraday_par})\\
    \hline
    Real & 4.27 & 0.79 & 1.09 & 0.16 & 0.43 & -0.64 & -0.97 & 0.69 & -0.98 \\
    Imaginary  & -1.47 & 0.25 & -0.31 & 0.18 & 0.41 & -1.1 & & & \\
    Magnitude & 4.51 & 0.83 & 1.13 & 0.24 & 0.59 & 1.3 & & &  \\
   \end{tabular}
   \end{center}
  \caption{Dimensionless complex WFT coefficients for the  fluctuating electric field $\mathbf{E}'$ and  magnetic field $\mathbf{B}'$ for the local plane-wave mode  $(k_{\perp 1}d_{i,0},k_{\perp 2}d_{i,0}, k_{||}d_{i,0}) = (1.97,0.00,0.34)$ , along with determinations of the real frequency $\omega$ from each of equations (\ref{eq:faraday1}), (\ref{eq:faraday2}), and (\ref{eq:faraday_par}) normalized to the upstream cyclotron frequency, $\Omega_{i,0}$.}
  \label{tab:wvmdtable}
\end{table}

%----------------------------------------------------------------
\subsection{Comparison to Linear Wave Modes}
\label{sec:comp2lin}

To determine the linear wave mode associated with the dominant wave vector identified through our WFT analysis at $x/d_{i,0}=39.875$, we will compare the real frequency computed from Faraday's Law to solutions for the frequency from the Vlasov-Maxwell linear dispersion relation for that wave vector using the PLUME dispersion relation solver \citep{klein2015predicted}.  To do so, we must employ the plasma conditions locally at $x/d_{i,0}=39.875$ in the frame of the plasma.  Relative to the upstream values of magnetic field, density, and ion and electron temperature, the local parameters have values  $B/B_0=1.709$, $n_i/n_0=n_e/n_0=2.045$, $T_i/T_{i0}=3.510$, and $T_e/T_{e0}=1.611$. Computing the local dimensionless parameters yields $\beta_i=2.240$, $T_i/T_e=2.179$, $k_\parallel d^{(loc)}_i=0.2378$, and $k_{\perp 1} d^{(loc)}_i=1.371$, where the ratio of the local ion inertial length to its upstream value is $d^{(loc)}_i/d_{i,0}=(n_0/n_i)^{1/2}=0.699$.
Similarly, the frequency from \eqref{eq:faraday2} must be converted to a normalization relative to the local ion cyclotron frequency $\Omega^{(loc)}_i/\Omega_{i,0} =(B/B_0)$, yielding a wave  frequency $\omega/\Omega^{(loc)}_i=0.404$.

\begin{figure} 
    \centering
    \includegraphics[width=1.\textwidth]{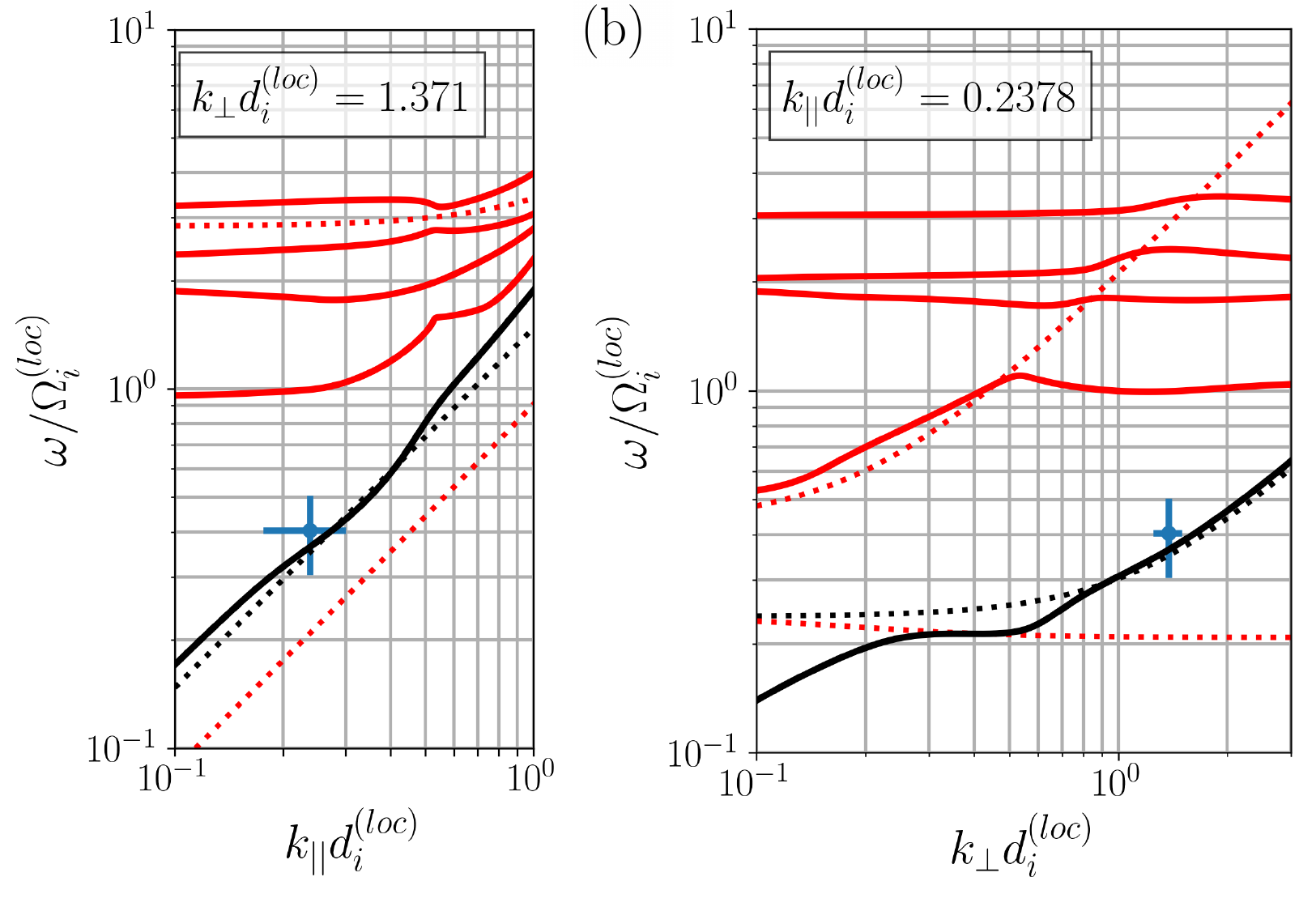}
    \caption{Comparisons between the measured frequency and wavelength of the ripple on the surface of the shock using empirical dispersion relations \citep{{klein2012using,howes2014validity}} (dotted lines), and dispersion relation for using PLUME, a Vlasov-Maxwell linear dispersion relation solver \citep{klein2015predicted} (solid lines). Blue points are measured wavelength and frequency of the dominant wavemode. The plotted dispersion relations are kinetic Alfv\`en wave (black), fast magnetosonic wave (red, lowest frequency), and first three ion Bernstein modes (red, higher frequencies), and slow magnetosonic wave (green). Frequency is normalized to the local ion cyclotron frequency $\Omega^{(loc)}_i$ and wavelength is normalized to the local ion inertial length $d^{(loc)}_i$. \emph{Left:} Dispersion relation as a function of $k_{||} d^{(loc)}_i$ at fixed $k_{\perp} d^{(loc)}_i= 1.371$. \emph{Right:} Dispersion relation as a function of $k_{\perp} d^{(loc)}_i$ at fixed $k_{||} d^{(loc)}_i = 0.2378$.}
    \label{fig:kawsweeps}
\end{figure}

The solutions from the PLUME solver for scans over $k_{||} d^{(loc)}_i$ and $k_{\perp 1} d^{(loc)}_i$ are presented in Figure \ref{fig:kawsweeps}. We compare these PLUME solutions to empirical dispersion relations from \citet{klein2012using} and \citet{howes2014validity} for a kinetic Alfv\'en wave valid in the limit $k_\parallel d_i \ll 1$,
\begin{equation}
    \frac{\omega}{\Omega_i}=k_{||}d_i\sqrt{1+\frac{(k_{\perp} d_i)^2}{1+2/\beta_i(1+T_e/T_i)}},
    \label{eq:kaw}
\end{equation}
and from the MHD dispersion relation for the fast and slow magnetosonic waves valid in the limit $k d_i \ll 1$, 
\begin{equation}
    \frac{\omega}{\Omega_i} = k d_i \Bigg[\frac{1+\beta_i(1+T_e/T_i)\pm\sqrt{[1+\beta_i(1+T_e/T_i)]^2-4\beta_i (1+T_e/T_i)(k_{||}^2/k^2)}}{2} \Bigg], 
    \label{eq:fast}
\end{equation} 
where the upper (lower) sign corresponds to the fast (slow) magnetosonic wave.  

In Figure~\ref{fig:kawsweeps}(a) and (b), we plot the PLUME solutions for the fast magnetosonic wave (lowest frequency solid red curve) along with the three lowest ion Bernstein modes (the $n=1$ through $n=3$ modes are the successively higher frequency solid red lines). In panel (b), the MHD fast wave dispersion relation \eqref{eq:fast} (dotted red) agrees well with the fast magnetosonic wave for up to $k_\perp d^{(loc)}_i=0.5$ and $\omega/\Omega^{(loc)}_i=1$, at which point the fast magnetosonic wave undergoes a mode conversion to the $n=1$ ion Bernstein mode.  Note that, as $k_\perp d^{(loc)}_i$ increases, the fluid solution for the fast magnetosonic wave  \eqref{eq:fast} corresponds to the regions of increasingly higher $n$ ion Bernstein modes where those modes have a positive perpendicular group velocity $\partial \omega/\partial k_\perp$, as has been found in previous kinetic studies of the fast mode at nearly perpendicular propagation \citep{Swanson:1989,stix1992waves,li2001damping,Howes:2009b}.
We also plot the PLUME solutions for the kinetic Alfv\`en wave (solid black), which show good agreement with the empirical kinetic Alfv\`en wave solution equation \ref{eq:kaw} (dotted black), and we plot the MHD solution for the slow magnetosonic wave (dotted green).

\begin{figure}
	\centering
	\includegraphics[width=1\textwidth]{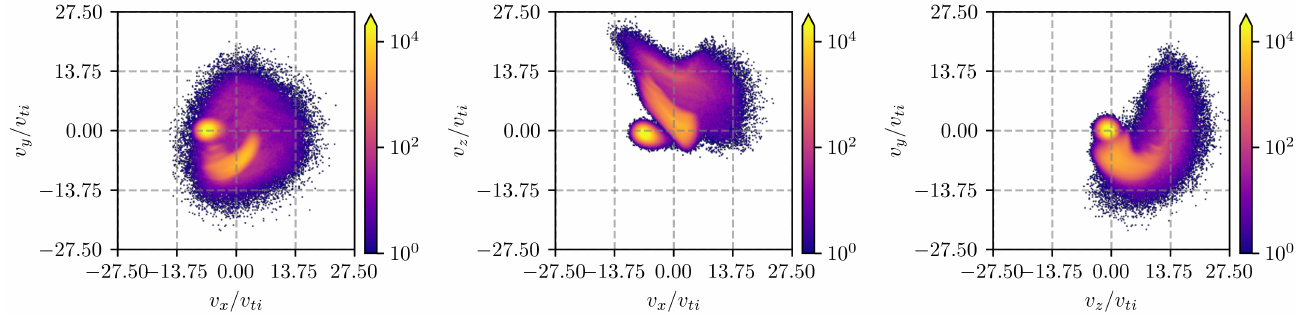}
	\caption{Distribution of ions in the ramp ($x/d_{i,0} = 39.875$) of a $M_A \approx 7.88$ and $\theta_{B_n} = 45\degree$ from a three-dimensional \texttt{dHybridR} simulation. The projection onto the $v_x$,$v_z$ (\emph{Middle}) shows three distinct populations of ions: the stream on the bottom, the population of having been reflected once in the middle, and the doubly reflected ions on top.}
	\label{fig:rampdist}
\end{figure}

To identify the wave mode responsible for the rippling of the shock front, we plot on Figure~\ref{fig:kawsweeps} the wave frequency $\omega/\Omega^{(loc)}_i=0.404$ calculated from \eqref{eq:faraday2} (blue point).  It is clear from this comparison that the Alfv\`en mode agrees with our frequency computation from the dominant unstable wave vector. We argue that this Alfv\'enic nature of the unstable fluctuation identifies the instability responsible for the shock ripple as the Alfv\'en ion-cyclotron instability (AIC) for four reasons. First, the AIC instability launches waves on the Alfv\'en branch (Alfv\'en waves, kinetic Alfv\'en waves, ion cyclotron waves) \citep{tajima1977Alfven}, which we have just shown are present in this simulation. 
Second, Figure \ref{fig:rampdist} shows that we have a partial ring distribution, which matches the form of a perturbed distribution that was used to derive the AIC instability in \cite{otani1988alfven}. Third, Figure \ref{fig:rampdist} also shows temperature anisotropy $T_{\perp}/T_{||}>1$ in the incoming beam (if a Bi-Maxwellian distribution is fit to the multiple populations present) that is required for the AIC instability. This result is consistent with previous studies \citep{winske1988magnetic,mckean1995wave,burgess2016microstructure}. Fourth, our measured temperature anisotropy, $T_{\perp}/T_{||}=2.01$, is above the marginal stability threshold established in \cite{Hellinger:2006}, $T_{\perp}/T_{||}>1.67$, for  the Alfv\`en ion cyclotron instability.

\begin{figure} 
    \centering
    \includegraphics[width=.99\textwidth]{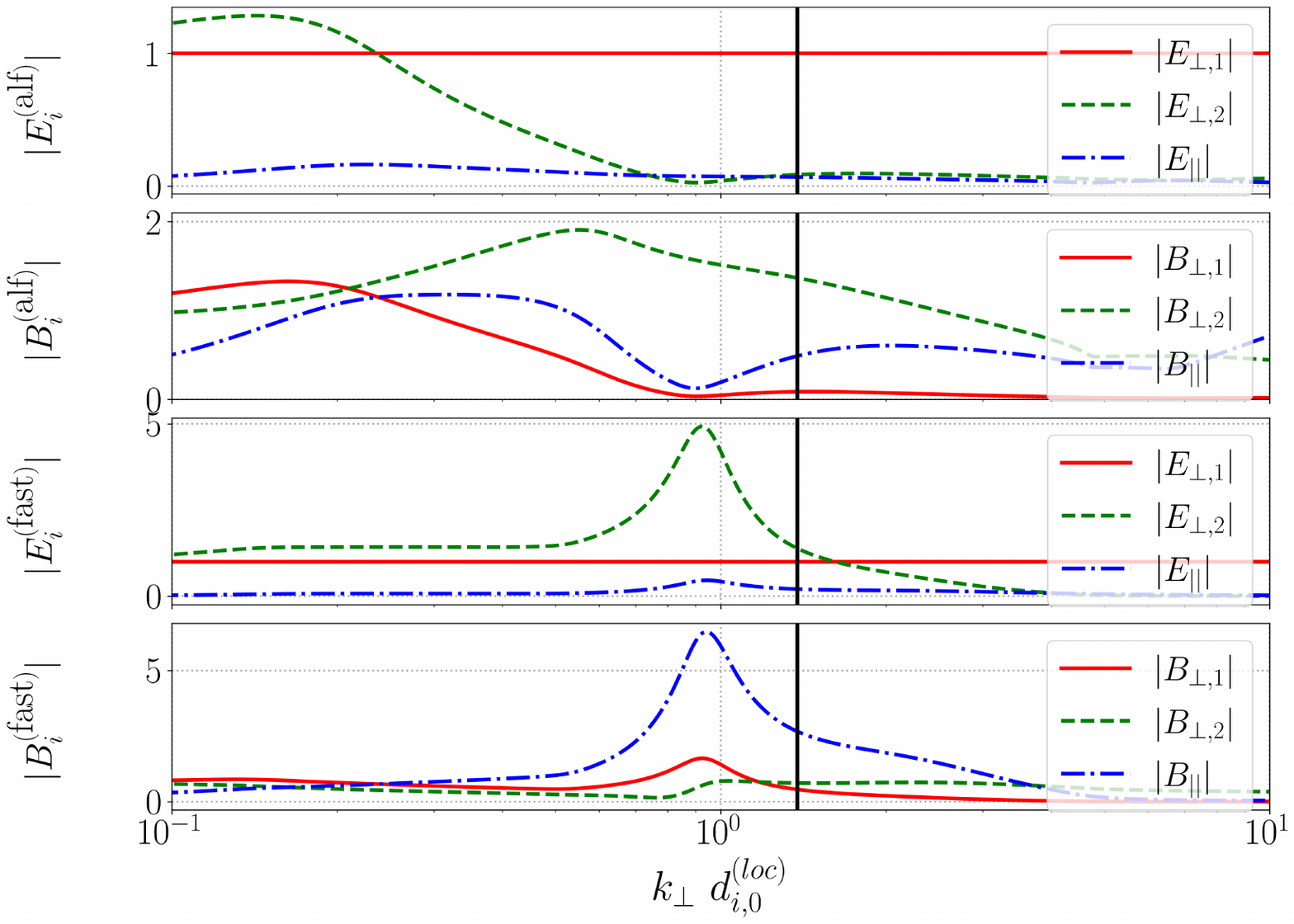}
    \caption{Magnitude of the components of the eigenfunction response of a kinetic Alfv\`en wave (\emph{Top Two}) and kinetic fast magnetosonic wave for the electric and magnetic fields (\emph{Bottom Two}) in a homogeneous plasma computed using PLUME, normalized to $|E_{\perp,1}|$, using local parameters at $x/d_{i,0} = 39.875$, as discussed in Section \ref{sec:comp2lin}. The vertical black line corresponds to the perpendicular wavenumber of the dominant wavemode discussed in Section \ref{sec:comp2lin}.}
    \label{fig:eigencomp}
\end{figure}

One should note, computing $\omega$ with \eqref{eq:faraday1}, \eqref{eq:faraday2}, and \eqref{eq:faraday_par} using measurements of $\mathbf{E}^\prime$ and $\mathbf{B}^\prime$ at some fixed $\mathbf{k}$ will not necessarily give similar values of $\omega$. This can occur as there may be multiple wave modes with different polarization may be superimposed. Figure \ref{fig:eigencomp} shows the polarization of a kinetic Alfv\'en wave and kinetic fast magnetosonic wave using the same parameters as those measured at $x/d_{i,0} = 39.875$ in the simulation. For $(k_{\perp 1} d^{(loc)}_i,k_{\perp 2} d^{(loc)}_i ,k_{||} d^{(loc)}_i)= (1.37,0.00,0.238)$, the amplitudes of $E_{||}$ and $E_{\perp 2}$ are very small relative to $E_{\perp 1}$ for the kinetic Alfv\'en wave, suggesting that \eqref{eq:faraday2} will give the best estimate of the wave frequency, while \eqref{eq:faraday1} and \eqref{eq:faraday_par} are worse choices for computing the frequency of an Alfv\'en wave with this wave vector. Thus, we choose the frequency computed using \eqref{eq:faraday2}, $\omega = 0.69 \Omega_i$, as the frequency of the dominant wavemode underlying the rippling of the shock surface.
 
We close with a brief discussion of the validity of using the Vlasov-Maxwell linear dispersion relation, derived for spatially homogeneous plasma conditions with no bulk plasma flow, to interpret the results of kinetic numerical simulations of a collisionless shock in which the plasma is strongly spatially inhomogeneous with a rapid flow of the plasma through the approximately steady-state structure of the shock transition.  The study of how kinetic instabilities arise and saturate in the presence of other dynamical plasma processes---such as plasma turbulence, magnetic reconnection, and collisionless shocks---represents an important area of research at the frontier of plasma physics.  For example, a recent analysis of spacecraft measurements from the \emph{Solar Orbiter} mission \citep{Muller:2020} argues that the complex interaction between turbulent fluctuations and kinetic instabilities ultimately regulates the proton-scale energetics of the solar wind \citep{Opie:2023}.  In the case of the collisionless shock investigated here, we identify the wavevector of the dominant fluctuation in the shock-rest frame, and then use Faraday's law to calculate the frequency of the plasma response in that frame of reference.  It is important the emphasize that, in the shock-rest frame, the non-Maxwellian ion velocity distributions shown in \figref{fig:rampdist} are relatively steady in time, and therefore it seems plausible that a sufficiently rapidly growing instability (with a growth rate of order the ion cyclotron period) will be able to arise locally from the free energy in the unstable ion velocity distribution.  Even in the presence of gradients of the plasma density, plasma temperature, and magnetic field, we still expect the linear response of the plasma to give rise to the wave-like fluctuations appropriate for the plasma parameters at that position.  An unanswered question, that is beyond the scope of this investigation, is how the super-Alfv\'enic flow of the plasma through shock transition competes with the growth of a kinetic instability, but the kinetic numerical simulation results presented here clearly show that, for the shock parameters $M_A \simeq 7.88$ and $\theta_{Bn}=45^\circ$, a kinetic instability is indeed able to give rise to shock rippling even in the presence of the bulk flow.

% Threshold used in klein2015
% b_i_par = .11
% 1+0.43/(b_i_par-0.0004)**.42=2.089

\section{Particle Energization}
\label{sec:partenerg}

The field-particle correlation technique allows us to understand the energetics in the 3D-3V phase of a kinetic plasma by analyzing the transfer of energy between fields and particles \citep{klein2016measuring,Howes:2017a,howes2017prospectus,klein2017diagnosing}. It has been used successfully to  explore energization of particles in kinetic turbulence simulations \citep{klein2017diagnosing,Howes:2018a,TCLi:2019,Klein:2020,Horvath:2020} and in \textit{in-situ} with spacecraft \citep{chen2019evidence,Afshari:2021}, as well as electron energization in simulations of collisionless magnetic reconnection \citep{mccubbin2022characterizing}, simulations of 1D-2V perpendicular collisionless shocks \citep{juno2021field}, and electron acceleration by Alfv\'en waves in the laboratory plasmas \citep{schroeder2021laboratory}. 

The rate of change of \emph{phase-space energy density} of species $s$, $w_s(\mathbf{r},\mathbf{v},t) \equiv m_s v^2 f_s(\mathbf{r},\mathbf{v},t)/2$, in a collisionless plasma can be determined by multiplying the Vlasov equation by $m_s v^2/2$, yielding
\begin{equation}
 \frac{\partial w_s}{\partial t} =
   - \mathbf{v}\cdot \nabla  w_s  -
  q_s\frac{v^2}{2}  \mathbf{E} \cdot \frac{\partial f_s}{\partial \mathbf{v}}
  - \frac{q_s}{c}\frac{v^2}{2} \left(\mathbf{v} \times \mathbf{B}\right)
      \cdot \frac{\partial f_s}{\partial \mathbf{v}}.
      \label{eq:dwdt}
\end{equation}
On the right-hand side of \eqref{eq:dwdt}, only the middle term with the electric field leads to a net change of energy of the particles \citep{klein2016measuring,Howes:2017a,klein2017diagnosing}, so we define the field-particle correlation by
\begin{equation}
C_{\mathbf{E}} (\mathbf{v},t) \equiv \bigg < -q_s \frac{v^2}{2} \frac{\partial f_s (\mathbf{r},\mathbf{v},t)}{\partial \mathbf{v}} \cdot \mathbf{E}(\mathbf{r},t) \bigg>,
\label{eq:fpc}
\end{equation}
where the angle brackets denote that the correlation is computed using either an average over a correlation interval in time or an integration over a volume in space. Note that integration of the correlation over velocity space of \eqref{eq:fpc} yields the rate of work done on species $s$ by the electric field, $\int d^3\mathbf{v} \, \, C_{\mathbf{E}} (\mathbf{v}) = \langle \, \mathbf{j}_s \cdot \mathbf{E} \, \rangle$. Since we are interested in determining the impact of each particular field component on the energization of particles, we may split the equation by each directional component $j$, 
\begin{equation}
C_{E_j} (\mathbf{v}) \equiv \bigg < -q \frac{v_{j}^2}{2} \frac{\partial f (\mathbf{r},\mathbf{v},t)}{\partial v_j} E_j(\mathbf{r},t) \bigg>,
\label{eq:fpc_comp}
\end{equation}
where the other contributions to $v^2$ in \eqref{eq:fpc} contribute zero net energization upon integration over velocity space \citep{juno2021field}.

The field-particle correlation technique generates \textit{velocity-space signatures} that contain both quantitative and qualitative features used here to further understand particle energization in collisionless shocks. The technique can be employed to probe the interactions between the electric field and the different populations of particles, \emph{e.g.}, the incoming beam of particles or the reflected particles at a supercritical shock. Previous approaches to understand particle energization in weakly collisional plasmas involve either tracking individual particles in phase space \citep{caprioli2014simulations} or reducing the fundamental `3D-3V' behavior of a kinetic plasma to the fluid quantity, such as $\mathbf{j} \cdot \mathbf{E}$ \citep{Gershman:2017}. These previous approaches have notable limitations in developing a full understanding of the energy transport in phase space: by following either only individual particles or velocity-space-integrated quantities, it is difficult to assess the interactions between the fields and distinct populations of particles in velocity space.  The field-particle correlation technique is especially useful in supercritical shocks, where the effect of the electric field on the small population reflected particles may comprise a significant fraction of the particle energization at the shock.

To implement the field-particle correlation technique for use with simulation data from particle-in-cell (PIC) codes, a straight-forward approach would be to specify a small spatial volume and bin all of the particles within that volume into velocity-bins, generating a velocity distribution function within that volume $f(\mathbf{v})$.  One would then be able to take the velocity derivatives of that distribution function and correlate them with the components of the electric field, as given by (\ref{eq:fpc_comp}).  However, in this approach, the electric field must be averaged over the spatial integration volume, losing the spatial dependence of the electric field within the volume. A solution to this issue is to use an alternative formulation of the correlation at time $t_0$ \citep{chen2019evidence},
\begin{equation}
C^\prime_{E_j} (\mathbf{v}) = \bigg< q v_j f(\mathbf{r},\mathbf{v},t_0) E_j(\mathbf{r},t_0) \bigg>
\label{eq:binFPC}
\end{equation}
where the electric field determined at the location of each particle and the brackets indicate an integration over a spatial volume. For shock problems, we typically choose a volume over a small range $\Delta x$  in the direction normal to the shock and covering the full transverse extent of the domain $L_y \times L_z$. As presented in \citet{chen2019evidence}, this alternative correlation $C^\prime_{E_j}$
 can then be used to compute the standard correlation $C_{E_j}$ through the relation
\begin{equation}
C_{E_j} (\mathbf{v}) = -\frac{v_j}{2} \frac{\partial C^\prime_{E_j}(\mathbf{v})}{\partial v_j} + \frac{C^\prime_{E_j}(\mathbf{v})}{2}
\label{eq:cprimetoc}
\end{equation}
It is important that the correlation is computed this way to retain the variation of the electric field within the integration volume.   We elaborate on the full details of the procedure of computing the field-particle correlation using data from PIC codes in appendix \ref{appendix:FPCinPIC}.

To separate the contribution to the particle energization by the steady-state shock physics from the energization due to the instability physics, we separate the transverse-plane averaged contribution from the fluctuating contribution, as given by \eqref{eq:avgfields} and  \eqref{eq:fluctfields},  for both the electric field $\mathbf{E}(x,y,z)=\overline{\mathbf{E}}(x)+ \delta \mathbf{E}(x,y,z)$ and the ion velocity distribution function  $f(\mathbf{r},\mathbf{v})=\overline{f}(x,\mathbf{v})+ \delta f(\mathbf{r},\mathbf{v})$.  Substituting these decompositions into \eqref{eq:fpc_comp}, we obtain
\begin{eqnarray}
C_{E_j} (\mathbf{v})& =  &\left[ 
\left\langle -q \frac{v_{j}^2}{2}  \frac{\partial \overline{f} (x,\mathbf{v})}{\partial v_j} \overline{E}_j(x) \right\rangle
+ \left\langle -q \frac{v_{j}^2}{2}  \frac{\partial \overline{f} (x,\mathbf{v})}{\partial v_j} \delta E_j(\mathbf{r}) \right\rangle \right.  \nonumber \\
&  + &  \left. \left\langle -q \frac{v_{j}^2}{2}  \frac{\partial \, \delta f (\mathbf{r},\mathbf{v})}{\partial v_j} \overline{E}_j(x)  \right\rangle
+ \left\langle -q \frac{v_{j}^2}{2}  \frac{\partial \, \delta f (\mathbf{r},\mathbf{v})}{\partial v_j} \delta E_j(\mathbf{r}) \right\rangle \right].
\label{eq:separate}
\end{eqnarray}
If the spatial average indicated by the angle brackets (also indicated by line over variable) denotes integration over the full transverse extent of the domain in $(y,z)$, with periodic boundary conditions in that transverse plane, then the two middle terms on the right-hand side of \eqref{eq:separate} contribute nothing since $\left\langle \delta E_j(\mathbf{r}) \right\rangle=0$ and  $\left\langle \partial  \delta f (\mathbf{r},\mathbf{v})/\partial v_j  \right\rangle=0$ by the definitions \eqref{eq:avgfields} and \eqref{eq:fluctfields}.  Therefore, under the transverse-plane average, we obtain a clean separation of the steady-state shock physics from the instability physics, defining the \emph{steady-state shock energization} through the \emph{averaged correlation}, $\overline{C}_{E_j}$, given by
\begin{equation}
    \overline{C}_{E_j}(\mathbf{v}) = \bigg < -q \frac{v_{j}^2}{2} \frac{\partial \overline{f} (x,\mathbf{v})}{\partial v_j} \overline{\mathbf{E}}(x) \bigg>,
    \label{eq:cbar}
\end{equation}
and the \emph{instability energization} through the \emph{fluctuating correlation}, $\widetilde{C}_{E_j}$, given by 
\begin{equation}
    \widetilde{C}_{E_j}(\mathbf{v}) = \bigg < -q \frac{v_{j}^2}{2} \frac{\partial \, \delta f (\mathbf{r},\mathbf{v})}{\partial v_j} \delta E_j(\mathbf{r}) \bigg>.
    \label{eq:ctilde}
\end{equation}
Furthermore, when we integrate over the full transverse extent of the domain, we may simply use the full ion velocity distribution function  $f(\mathbf{r},\mathbf{v})$ in the derivative for both \eqref{eq:cbar} and \eqref{eq:ctilde}, since the difference vanishes upon the spatial integration.
Thus, the averaged correlation $\overline{C}_{E_j}(f)$ captures the transfer of energy between the particles and the averaged fields and the fluctuating correlation $\widetilde{C}_{E_j}(f)$ captures the transfer of energy between the particles and the fluctuating fields arising from instabilities. 

\subsection{Steady-State Shock Energization}
\begin{figure} 
    \centering
    \includegraphics[width=0.99\textwidth]{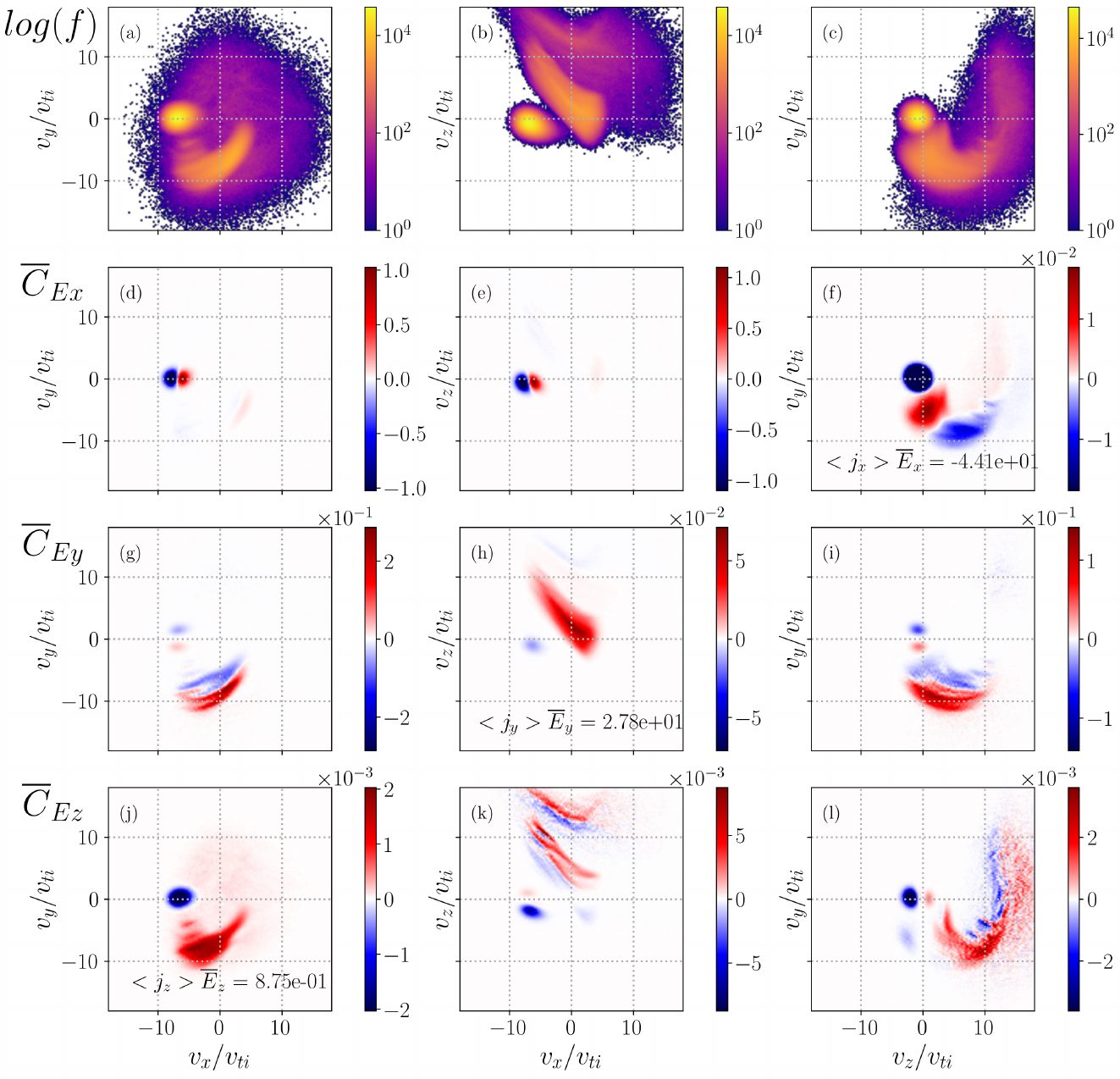}
    \caption{(a)--(c) Total velocity distribution function of ions $f_i(x,\mathbf{v})$ at the transition from the foot to the ramp of the shock at $x_0/d_{i,0} = 39.875$. (d)--(l) Average energization from the field-particle correlation $\overline{C}_{E_j}(x_0,\mathbf{v})$, where we integrate the 3V velocity-space over the third velocity-space coordinate in each column. All quantities are computed in the shock-rest frame and normalized to $n_0$, the upstream particle density.}
    \label{fig:cbarfull9pan}
\end{figure}

We present our field-particle correlation analysis of the rate of ion energization due to the steady-state shock dynamics using the averaged correlations $\overline{C}_{E_j}(\mathbf{v})$ for each component of the averaged electric field $\overline{E}_j$ in Figure~\ref{fig:cbarfull9pan}. The correlation is computed over an integration volume with normal thickness $\Delta x=0.25 \, d_{i,0}$, centered at $x_0/d_{i,0}=39.875$, and area covering the full transverse extent of the domain $L_y\times L_z=12 d_{i,0} \times 12 d_{i,0}$. We separate the energization by each of the three components of the averaged electric field in the simulation coordinates, $(\overline{E}_x,\overline{E}_y,\overline{E}_z)$, and present three views of the 3V velocity space using integrations along each of the three velocity space dimensions, for a total of nine panels.  For example, the reduction along the $v_z$ dimension for the $\overline{E}_y$ averaged correlation is given by $\overline{C}_{E_y}(v_x,v_y) = \int dv_z \overline{C}_{E_y}(v_x,v_y,v_z)$.  In addition, we plot the same three reduced views of the ion velocity distribution $f_i(\mathbf{v})$ in the top row, using a logarithmic color map to highlight the different small populations of reflected particles.  

Note that the \emph{net} energization by each of the three field components is most clearly visualized along the reverse diagonal of the nine-panel correlation plots, panels (f), (h), and (j). The dependence of the correlation \eqref{eq:cbar} on $\partial \overline{f}/\partial v_j$, combined with the fact that this derivative must pass through zero along the $v_j$ axis, implies that the correlation with $\overline{E}_j$ consists of positive and negative regions along the $v_j$ axis.   Using the red-white-blue colormap, it can be difficult to assess by eye the net rate of energization.  But upon the integration over the $v_j$ dimension---given by panels (f), (h), and (j) on the reverse diagonal---the net energization is easily observed.  The quantitative value of the net rate of energization by the $\overline{E}_j$ over the integration volume, $\langle j_{j,s} \overline{E}_j \rangle$,  is also plotted at the top of the reverse diagonal panels. 

Furthermore, to assist in interpretation of the velocity-space signatures, we note that when a bipolar signature is observed (consisting of adjacent blue and red regions), the weighting by $v_j^2$ in the correlation means that the color further from the origin (at higher velocity magnitude $|v_j|$) usually dominates. To standardize our terminology for bipolar signatures, we will state the color with the lower magnitude of velocity $|v_j|$ first.  Therefore, a blue-red bipolar signature generally denotes net particle energization, and a red-blue bipolar signature typically indicates a net loss of particle energy. One can interpret bipolar signatures as particles being accelerated/decelerated by the $E_j$ component of the electric field from the `blue' region to the `red' region in a collisionless system.

The contribution from averaged cross-shock electric field $\overline{E}_x$, evaluated at $x_0/d_{i,0}=39.875$, to the rate of change of phase-space energy density due to the steady-state shock energization is given by the velocity-space signatures in Figure~\ref{fig:cbarfull9pan}, panels (d)--(f). Panels (d) and (e) show the deceleration of the incoming ion beam in the $(v_x,v_y)$ and $(v_x,v_z)$ projections, with a red-blue bipolar signature indicating a net loss of energy by the incoming beam population. This net loss of ion energy by $\overline{E}_x$ is clearly shown in panel (f) as the dark blue region at $(v_z,v_y) \simeq (0,0)$.  Also visible in panel (f) is the small population of reflected ions heading back upstream ($v_x>0$) at $v_y/v_{ti} \simeq -5$ that gain energy from the cross-shock electric field, and a subsequent loss of energy by  $\overline{E}_x$ as those reflected particles return downstream at $v_y/v_{ti} \simeq -9$  and $ 3\lesssim v_z/v_{ti} \lesssim 7$.

The primary mechanism of ion acceleration at quasi-perpendicular collisionless shocks is mediated by the the motional electric field, $\overline{E}_y$, with the three viewpoints of $\overline{C}_{E_y}$ shown in panels (g)--(i).  In panel (g), the dominant velocity-space signature is the blue-red `crescent' indicating acceleration of the reflected ion population by the motional electric field \citep{juno2021field,Juno:2022}, a mechanism known as \emph{shock-drift acceleration} \citep{Paschmann:1982, Sckopke:1983, Anagnostopoulos:1994, Anagnostopoulos:1998, Ball:2001, Anagnostopoulos:2009, Park:2013}.  The same blue-red crescent signature is visible from a different viewpoint in panel (i).  The net ion energization due to shock-drift acceleration by the 
motional electric field is most easily viewed in panel (h), where the red diagonal feature is an additional velocity-space signature of shock-drift acceleration \citep{Juno:2022} and is perpendicular local magnetic field direction at this point $x/d_{i,0}=39.875$, where the shock transitions from the foot to the ramp.  Integration of $\overline{C}_{E_y}$ over velocity space shows that the net rate of positive ion energization, $\langle j_y \overline{E}_y \rangle$, is dominated by the acceleration of the reflected ion population by the averaged motional electric field.  
A key result of our field-particle correlation analysis of the steady-state shock energization is the \emph{velocity-space signature of shock-drift acceleration}, shown in panels (g)--(i), together indicating the acceleration of the reflected ion population by upstream the motional electric field for a $\theta_{Bn}=45^\circ$ shock.

The steady-state shock energization due to the self-consistently generated  $\overline{E}_z$ is shown in panels (j)--(l).  In
panel (k), we show the energization of three separate populations of ions: (i) a loss of energy for the incoming ion beam at $v_z \simeq 0$; (ii) a  small population of ions that have been reflected once in the range  $2 \lesssim v_z/v_{ti} \lesssim 12$;  and (iii) a smaller population of ions that have been reflected a second time from the shock (and may be escaping upstream, as discussed in \citet{Juno:2022}) at $12 \lesssim v_z/v_{ti} \lesssim 18$.
The net loss of energy by the incoming beam and net gain of energy by the two populations of reflected ions by $\overline{C}_{E_z}$ is clearly seen in panel (j).
Note, however, that the net rate of energization by $\overline{E}_z$ is more than a factor of 30 smaller than 
that by $\overline{E}_x$ and $\overline{E}_y$.

Using this analysis, we can analyze the energization of each population of particles separately. With the field-particle correlation technique, one can clearly separate the net loss of energy of the incoming ion beam from the energization of the reflected ion population. A fluid description of particle energization providing only $\mathbf{j}_i \cdot \mathbf{E}$, a quantity integrated over velocity space, loses this clear separation of energization mechanisms operating on distinct populations of ions in velocity space.

\subsection{Instability Energization}

\begin{figure} 
    \centering
    \includegraphics[width=0.99\textwidth]{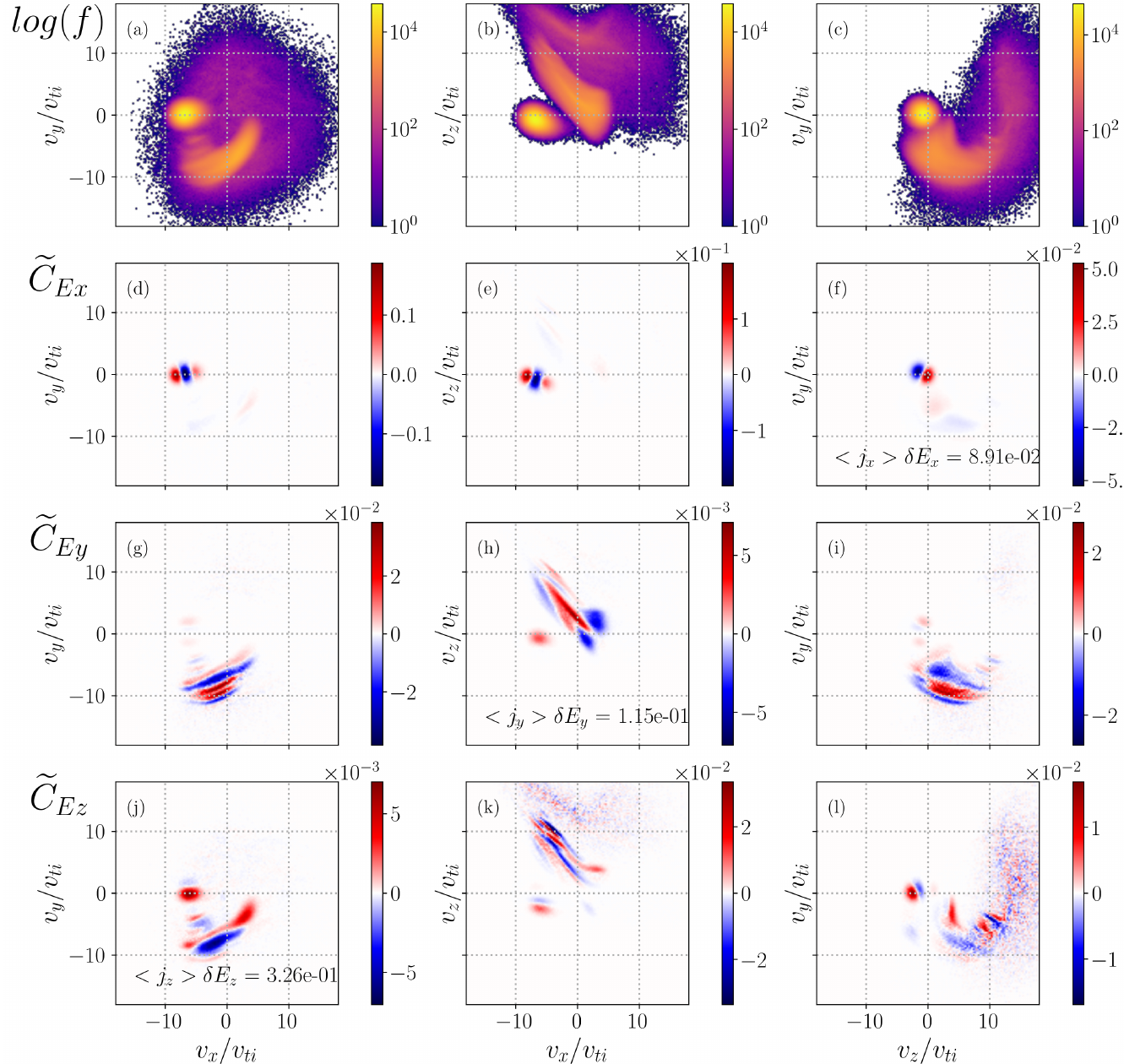}
    \caption{(a)--(c) Total velocity distribution function of ions $f_i(x,\mathbf{v})$ at the transition from the foot to the ramp of the shock at $x_0/d_{i,0} = 39.875$. (d)--(l) Instability energization from the fluctuating correlation $\widetilde{C}_{E_j}(x_0,\mathbf{v})$, where we integrate the 3V velocity-space over the third velocity-space coordinate in each column. All quantities are computed in shock-rest frame.}
    \label{fig:ctilde9pan}
\end{figure}

We present our field-particle correlation analysis of the rate of ion energization due to the kinetic instabilities driven at the shock using the fluctuating correlations  $\widetilde{C}_{E_j}$  for each component of the fluctuating electric field $\delta E_j$ in Figure~\ref{fig:ctilde9pan}.
These results are presented using the same format as used for  the steady-state shock energization in Figure \ref{fig:cbarfull9pan}, enabling direct comparisons of features and amplitudes of the energization rates.

First, we note that the net rates of energization due to the three fluctuating electric field components  $\delta E_j$ is more than two-orders of magnitude smaller than the rate of energization by the steady-state physics of the shock transition. Second, the rate of energy transfer from all three components of $\widetilde{C}_{E_j}$ is positive, meaning that the instability-generated fluctuations are being damped at this position, giving their energy to the particles.  In the next section, we will explore how the  net rate of energization by instabilities varies as a function of the normal distance through the shock.

The key new qualitative feature of interest in these fluctuations correlations $\widetilde{C}_{E_j}$ is the appearance of a `tripolar' signature, such as that arising from the fluctuating normal component of the electric field $\delta E_x$ in panels (d) and (e). We do not observe any such  tripolar features due to the steady-state shock physics in Figure \ref{fig:cbarfull9pan}.  Such a tripolar feature is a consequence on the non-uniformity of the electric field across the transverse domain, \emph{e.g.} shock ripple, as we explain below. This tripolar feature indicates increasing phase-space energy at velocities flanking the average velocity of the incoming stream. 

\begin{figure} 
    \centering
    \includegraphics[width=.69\textwidth]{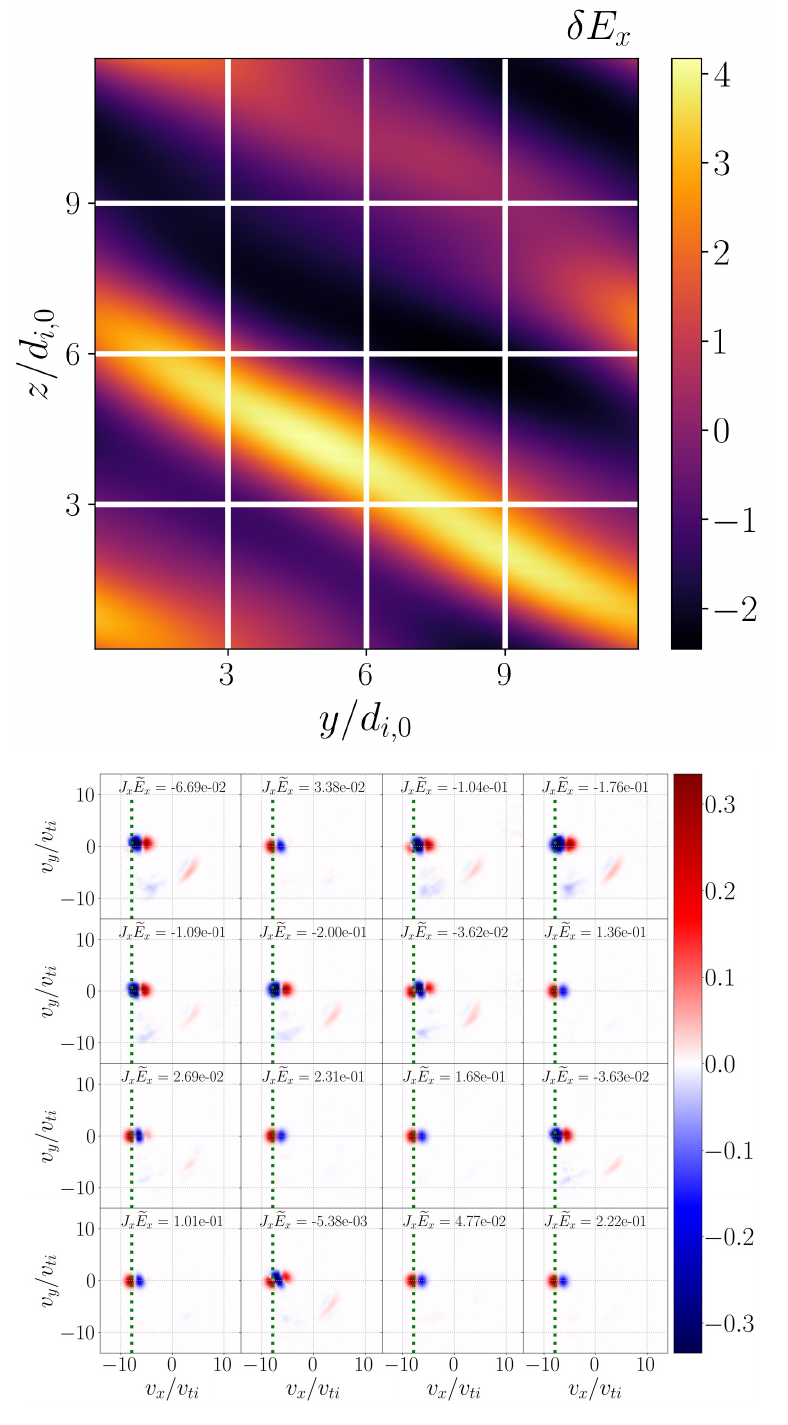}
    \caption{(\emph{Top:}) Fluctuating cross-shock electric field $\delta E_x$ at $x_0/d_{i,0} = 39.875$, plotted across the transverse plane, with subregion boundaries indicated (white lines). (\emph{Bottom:})
    Fluctuating correlation $\widetilde{C}_{E_x}(v_x,v_y)$, generated using spatial subregions of size $3 d_{i,0} \times 3 d_{i,0}$  in the $y,z$ plane, with the same range $\Delta x=0.25 d_{i,0}$ centered at $x_0/d_{i,0} = 39.875$ as used in Figures \ref{fig:cbarfull9pan} and \ref{fig:ctilde9pan}. The dotted green vertical line on each panel indicates the average velocity of the incoming ion stream in the shock-rest frame.}
    \label{fig:subboxFPCs}
\end{figure}

The tripolar feature in Figure \ref{fig:cbarfull9pan}(d) can be explained by computing the correlation using sub-regions across the transverse plane of the simulation. In the top panel of Figure \ref{fig:subboxFPCs}, we plot the normal component of the fluctuating electric field $\delta E_x$ across the full transverse plane $(y,z)$ at $x/d_{i,0}=39.875$. We then divide this plane up into 16 subregions of size $3 d_{i,0} \times 3 d_{i,0}$ (white lines) and compute the 
fluctuating correlation  $\widetilde{C}_{E_x}(v_x,v_y)$ in each of these sub-regions, plotted in the lower panel. The second column of the lower panel illustrates how the sum of each of these sub-regions combines to produce the tripolar signature.  Numbering the four elements of the second column from top to bottom, the first and third elements have $\delta E_x>0$ and yield a blue-red bipolar (positive energization) signature, while the second element has $\delta E_x<0$ and yields a red-blue bipolar (negative energization) signature. The fourth element has a mixture of  $\delta E_x>0$  and $\delta E_x<0$ within the sub-region, and yields a red-blue-red tripolar signature.  Critically, the $v_x$ value of the zero crossing of the bipolar signatures in the first three elements shifts slightly to right in first and third elements and slightly to the left in the second element. (The vertical dotted green line, at the average velocity of the incoming ion stream in the shock-rest frame, assists in visualization of this small shift.) Therefore, their sum would yield a net positive (red) value at the lowest $|v_x|$ values, a net negative (blue) at intermediate $|v_x|$ values, and a positive (red) value at the highest $|v_x|$ values---their sum would then be a red-blue-red tripolar signature, with a net positive rate of ion energization.  Thus, the qualitative appearance of a tripolar signature indicates energization by the fluctuating electric field (across the transverse plane) associated with an instability, and the velocity-space integrated rate of energization yields the net ion energization rate by that instability.

Using our separation of the steady-state shock physics from the instability physics, we find that the instability driving the rippling of the shock surface in the ramp accounts for $\lesssim 1$\% of the total ion energization. Despite this small amplitude of energy transfer, we are able to isolate the dominant wave mode causing the shock ripple, distinguish shock energization from instability energization using the velocity-space signatures,  and quantitatively analyze the rate  of energy transfer to the ions in velocity space using the field-particle correlation technique. With these methods, one can analyze the potentially different energization of distinct populations of particles, even for weak mechanisms.

\begin{figure} 
    \centering
    \includegraphics[width=.99\textwidth]{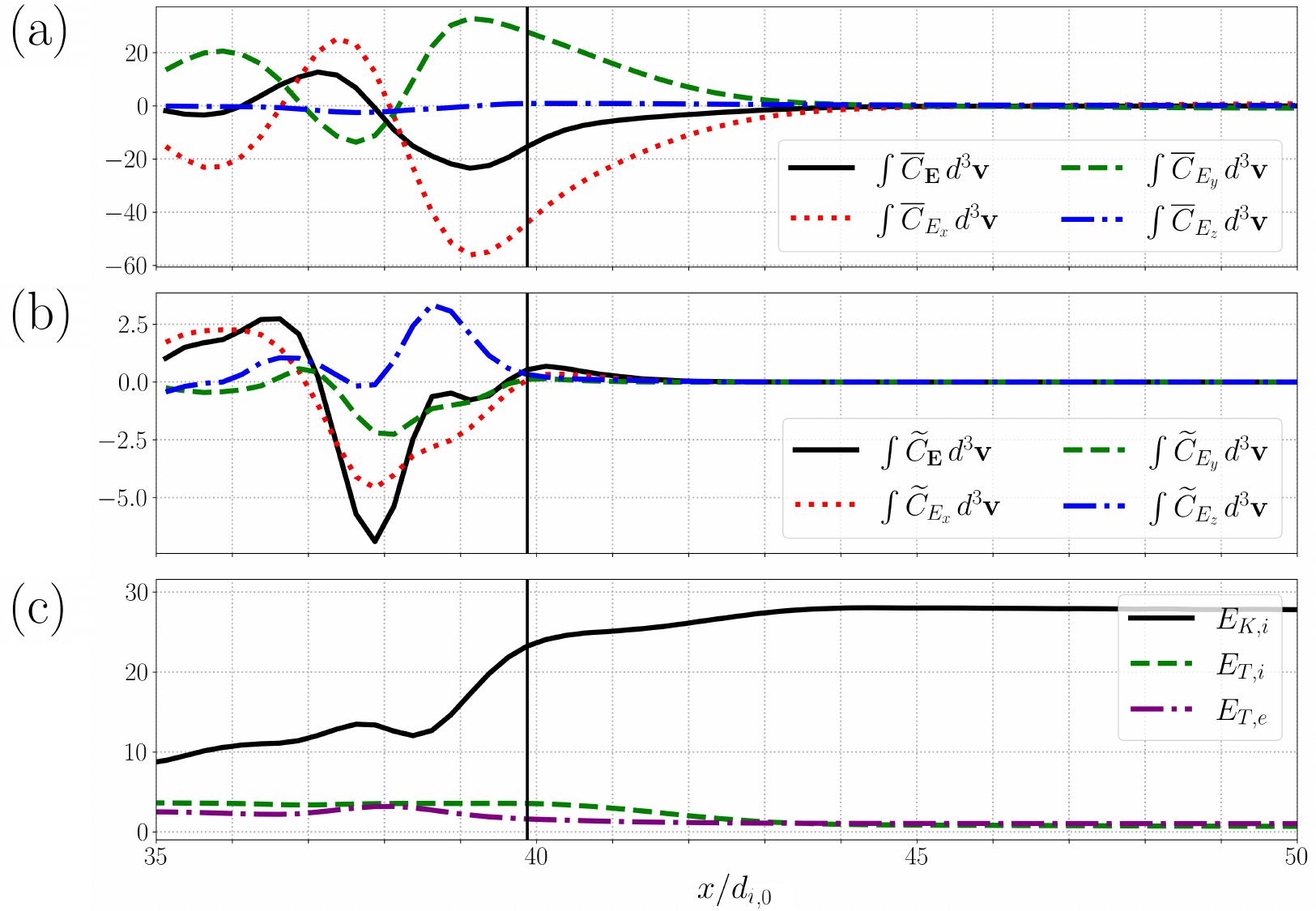}
    \caption{Comparison of energy transfer between the particles and transverse-plane averaged fields(\emph{Top Row}) as well as between that particles and the fluctuating fields(\emph{second row}), and kinetic energy/thermal energy of ions (\emph{Bottom Row}) near the shock transition region in the shock-rest frame. The vertical black line indicates the position that the instability isolation method was applied in Section \ref{sec:id}. Quantities are normalized as specified in Section \ref{sec:simsetup} and by $n_0$, the upstream particle density.}
    \label{fig:Corvsx}
\end{figure}

\subsection{Steady-State Shock and Instability Energization through the Shock}
\label{sec:ceivsx}

The profile of energization of ions throughout the shock due to the averaged fields and the fluctuating fields is shown in Figure \ref{fig:Corvsx}, where the total rate of energization can be computed from integrating the field-particle correlation over velocity space using
\begin{equation}
    \int d^3\mathbf{v} \, \,  C_{\mathbf{E}} (\mathbf{v})  = \mathbf{j} \cdot \mathbf{E} .
\end{equation}
Thus, we can compute energization due to the steady-state shock physics and to the instability physics separately  by computing $\int d^3\mathbf{v} \, \, \widetilde{C}_{\mathbf{E}} (\mathbf{v})$ and $\int d^3\mathbf{v} \, \, \overline{C}_{\mathbf{E}} (\mathbf{v}) $, respectively. 

The  $\int d^3\mathbf{v} \, \, \overline{C}_{E_x}(x) (\mathbf{v}) $ term  (top panel, red dotted) shows deceleration of the ions (loss of energy) by the cross shock electric field throughout the shock transition region, $44 \gtrsim x/d_{i,0} \gtrsim 38$.  We see net acceleration of ions (gain of energy) throughout the same shock transition region by $\overline{E}_y$ in the $\int d^3\mathbf{v} \, \, \overline{C}_{E_y}(x) (\mathbf{v}) $ term (top panel, green dashed), which accelerates the reflected ions through the shock-drift acceleration mechanism, as demonstrated by our field-particle correlation analysis in Figure~\ref{fig:cbarfull9pan}(h). The work done by the $\overline{E}_z$ component of the electric field on the ions is generally negligible relative to that due to the other components.

The relative rate of energization of the ions due to the instability-driven fluctuating fields is small compared to the steady-state shock energization throughout the entirety of the shock transition region, with amplitudes often much less than 10\% of the steady-state shock energization. 
The ion kinetic instabilities that generate the electromagnetic field fluctuations observed in this simulation arise by tapping free energy in unstable ion velocity distributions caused by the steady-state shock physics in the shock transition. Therefore, the condition that the velocity-integrated total fluctuating correlation is negative, $\int d^3\mathbf{v} \, \, \widetilde{C}_{\mathbf{E}}(\mathbf{v})<0 $, indicates clearly the regions where these instabilities arise. As shown in Figure~\ref{fig:Corvsx}(b), the ions drive these instabilities over the range $37 \lesssim x/d_{i,0} \lesssim 39.5$, which corresponds to the ramp and overshoot regions of the shock.  These instability-driven fluctuations appear to propagate a short distance upstream into the foot of the shock, before they are fully damped out upon reaching $x/d_{i,0} \gtrsim 42$.  Thus, at the position $x_0/d_{i,0}=39.875$ analyzed in our field-particle correlation analysis, the unstable electromagnetic fluctuations are being damped and leading to a small, but positive, net energization of the ions at that position, as shown in Figure~\ref{fig:ctilde9pan}.  Unstable electromagnetic fluctuations may also be swept downstream to $x/d_{i,0} \lesssim 37$ with the bulk plasma flow, appearing as turbulent fluctuations downstream that we expect will eventually be damped and ultimately lead to heating of the downstream plasma.

\section{Conclusion}
\label{sec:conc}

We present an in-depth analysis of the shock ripple produced by a kinetic instability in a three-dimensional, hybrid PIC simulation of an oblique, supercritical shock.
We introduce a new approach---the instability isolation method---to separate the steady-state shock
physics from the physics of kinetic instabilities that may arise in the shock transition.  An extension of the field-particle correlation technique enables us to separate the impact of the shock physics on ion energization from that due to the instabilities.

The instability isolation method divides the variation along the shock normal direction due to steady-state shock
physics from the variation transverse to the shock normal due to kinetic instabilities. The electromagnetic fields are separated into the averaged fields and the fluctuating fields. A Wavelet-Fourier transform (WFT) is used to identify the dominant local wave vectors arising from the instability that lead to rippling of the shock front.  We apply Faraday's Law using this wave vector and its associated electric and magnetic field components to estimate the real frequency of the unstable mode. Comparing this frequency to the eigenfrequencies of the Vlasov-Maxwell linear dispersion relation demonstrates the Alfv\'enic nature of the fluctuations causing the shock ripple, likely arising from the Alfv\'en-ion cyclotron (AIC) instability driven by a perpendicular ion temperature anisotropy caused by compression of the plasma through the shock, in agreement with the findings of earlier studies \citep{winske1988magnetic,mckean1995wave,burgess2007shock,burgess2016microstructure}. 

The field-particle correlation technique uses single-point measurements of the electromagnetic fields and particle velocity distributions to obtain a velocity-space signature characteristic of a given particle energization mechanism and to compute the rate of energy transfer between the fields and the particles.  We describe a specific procedure for the implementation of this technique with data from particle-in-cell (PIC) based kinetic simulation codes.  Further, we describe a new extension of the technique to use separately the transverse-plane averaged fields to compute the steady-state shock energization through a newly defined averaged correlation \eqref{eq:cbar} and the fluctuating fields to compute the 
instability energization through a newly defined fluctuating correlation \eqref{eq:ctilde}.

The results of the averaged field-particle correlation, applied to measurements at the transition from the ramp to the foot of the shock, are presented in \figref{fig:cbarfull9pan}, with nine panels that present the energization due to the three components of the averaged electric field, each reduced along each of the three dimensions of 3V velocity space.  The correlation with the averaged motional electric field $\overline{C}_{E_y}(\mathbf{v})$, shown in panels (g)--(i), displays the characteristic velocity-space signature of shock-drift acceleration of reflected ions at an oblique shock, dominating the total ion energization in the shock foot.

Computation of the fluctuating correlation, shown in \figref{fig:ctilde9pan}, shows that the fluctuations causing the shock ripple---although they produce a measurable impact on the surface of the shock as seen here in simulation and in \textit{in-situ} observations \citep{johlander2016rippled,johlander2018shock}---have a negligible impact on the ion energization, with energization rates more than two-orders of magnitude smaller than the steady-state shock physics. But the fluctuating correlation does produce a new, qualitatively distinct `tri-polar' signature of ion energization, as seen in panel (d).  By decomposing the transverse plane into sub-regions, we demonstrate that the tri-polar signature arises from variations of the fluctuating electric field and fluctuating velocity distribution across the integration volume used to compute the distribution.  Therefore, the appearance of a tri-polar signature is indicative of energization by the electric field arising from an instability, leading to a qualitative means to identify particle energization by instabilities.

Integrating the averaged and fluctuating correlations over velocity-space yields the net ion energization at the spatial point of analysis.  This enables us to produce a profile of the energy transfer between the fields and the ions as a function of the normal distance through the shock, presented in \figref{fig:Corvsx}.  This shows that the instability experiences net wave growth in the ramp and overshoot regions of the shock, with a fraction of the wave energy propagating a short distance upstream into the foot of the shock before being entirely damped. The remaining instability-driven wave energy appears to propagate downstream with the bulk plasma flow, generating the oft-observed turbulence in the downstream plasma. 

Here we have developed new tools to characterize more completely the energetics of collisionless shocks in phase space and have shown their value for analyzing instabilities arising in shocks and the resulting impact of the instability-driven fluctuations on particle energization. 
Our success comparing linear kinetic theory predictions to analyzed simulation results shows that the instability isolation method can be used quantitatively to characterize the nature of instability-driven fluctuations, even for mechanisms that transfer small amounts of energy in the system.  Future work will apply this technique to shocks with higher Mach number in which instabilities are believed to have a more significant impact on the particle energization, in particular controlling the partitioning of upstream kinetic energy between ions and electrons, \emph{e.g.}, instances of non-adiabatic electron heating often observed in sufficiently high Mach number shocks.

\section*{Acknowledgements}
The authors thank Lynn B. Wilson for helpful discussion and advice during this project.

\section*{Funding}
Resources supporting this work were provided by the NASA High-End Computing (HEC) Program through the NASA Advanced Supercomputing (NAS) Division at Ames Research Center.
C.R.B., G.G.H., C.C.H., and S.C.  were supported by NASA grant 80NSSC20K1273.  G.G.H. was also supported by subcontract M99016CAC from the Southwest Research Institute.

\section*{Declaration of Interest}
The authors report no conflict of interest.

\section*{Data Availability Statement}
The data that support the findings of this study are openly available in Zenodo at \href{https://doi.org/10.5281/zenodo.7901521}{https://doi.org/10.5281/zenodo.7901521}.

\section*{ORCID iDs}

\noindent Collin Brown  \href{https://orcid.org/0000-0001-6113-4860}{https://orcid.org/0000-0001-6113-4860}\\
\noindent James Juno \href{https://orcid.org/0000-0001-6835-273X}{https://orcid.org/0000-0001-6835-273X}\\
\noindent Gregory G. Howes \href{https://orcid.org/0000-0003-1749-2665}{https://orcid.org/0000-0003-1749-2665}\\
\noindent Colby C. Haggerty \href{https://orcid.org/0000-0002-2160-7288}{https://orcid.org/0000-0002-2160-7288}\\
\noindent Sage Constantinou \href{https://orcid.org/0000-0002-8937-5620}{https://orcid.org/0000-0002-8937-5620}\\

\bibliographystyle{jpp}
\bibliography{bibs/abbrev.bib,bibs/corrugation_ripple.bib,bibs/fpc.bib,bibs/plume.bib,bibs/shocks_overview.bib,bibs/shocks_sims.bib,bibs/wlt.bib,bibs/shock_reformation.bib,bibs/random.bib}

\appendix
\section{Field-Particle Correlation in PIC Codes}\label{appendix:FPCinPIC}
The hybrid kinetic ion/fluid electron simulation code \texttt{dHybridR} \citep{haggerty2019dhybridr} represents the ion velocity distribution using the particle-in-cell (PIC) method. Therefore, the ion velocity distribution function at a given spatial position must be constructed by creating a histogram in 3V velocity space of all $N$ particles within a finite binning volume $\Delta V$ in configuration space.  This procedure yields an ion velocity distribution function with a noise level that scales as $N^{-1/2}$.  To minimize the noise, one may either run a PIC simulation with a large number of particles per cell, which is computationally costly, or choose a sufficiently large binning volume $\Delta V$ such that the number of particles $N$ within that volume is sufficiently large to yield a low level of noise. If the correlation is computed by first binning the ions into a velocity distribution function $f_i(\V{r},\mathbf{v})$ and then combining with the electric field, the average of the electric field over the binning volume must be used in the correlation. But for a sufficiently large binning volume needed to obtain a low-noise velocity distribution, the volume-averaged electric field obviously does not capture variations of the electric field within the binning volume, potentially leading to inaccurate results.

Below is a general procedure for computing the standard field-particle correlation  $C_{\V{E}}(\V{r},\V{v},t)$ using particle-in-cell code data with a total of $N$ particles throughout the full simulation domain volume $V$.  This procedure is designed to retain the variations of the electric field $\V{E}(\V{r},t)$ within the subvolume $\Delta V$ over which the distribution is computed.

\begin{enumerate}
\item  First, we divide the entire simulation domain into subvolumes $\Delta V_{ijk}$, where each subvolume is centered at position $(x_i,y_j,z_k)$ and the indices $(i,j,k)$ indicate the spatial position of the subvolume.  The total number of particles in each subvolume is  $N^{ijk}$, such that $N=\sum_{i,j,k} N^{ijk}$.
  
\item We create a uniform grid of velocity bins in 3V velocity space with linear velocity bin size $\Delta v$.  Each 3V velocity bin  $\Delta \mathcal{V}_{lmn}$ is centered at $(v_{x,l},v_{y,m},v_{z,n})$, where the indices $(l,m,n)$ indicate the velocity bin.  Therefore, the particles within a given velocity bin have a velocity with an $x$-component within the range $v_{x,l} - \Delta v/2 \le v_x \le v_{x,l} + \Delta v/2$, and
  equivalently for the $v_y$ and $v_z$ components.   The total number of particles in each velocity bin is  $N^{ijk}_{lmn}$, such that $N^{ijk}=\sum_{l,m,n} N^{ijk}_{lmn}$. In practice, we construct a finite number of velocity bins, covering from $-v_{max}$ to $v_{max}$ in  each velocity dimension, although the value of  $v_{max}$ can be changed based on the particular case of interest, \emph{e.g.}, higher Mach number shocks will require a larger  $v_{max}$ to capture the significant features of the particle velocity distributions.

\item For all of the particles $\alpha$ falling within a spatial
  subvolume $\V{r}_\alpha \in\Delta V_{ijk}$ and within the velocity bin
  $\V{v}_\alpha \in \Delta \mathcal{V}_{lmn}$, denoted by $\alpha \in \Delta V
  \Delta \mathcal{V}$ where $\alpha$ is the particle index, we compute
  the contribution for  each particle to the work done by the $\mu$th
  component of the electric field $E_\mu$, given by
\begin{equation}
  c'_{E_\mu,\alpha} = W_\alpha  q_\alpha  v_{\mu,\alpha}  E_{\mu}(\V{r}_\alpha).
\end{equation}
Here $W_\alpha$ denotes the weight of each PIC ``macroparticle'' $\alpha$, and the electric field is evaluated at the particle position, $ \V{E}(\V{r}_\alpha)$, computed using  tri-linear interpolation from the electromagnetic field grid.

\item The alternative field-particle correlation is computed at the subvolume center
  $\V{r}_{ijk}$ and velocity bin center $\V{v}_{lmn}$ by summing   $c'_{E_\mu,\alpha}$ over  all $N^{ijk}_{lmn}$ paticles within the 3D-3V phase-space volume,
  \begin{equation}
  C'_{E_\mu}(\V{r}_{ijk},\V{v}_{lmn})= \frac{1}{\Delta V \Delta \mathcal{V}}
 \sum_{\alpha  \in \Delta V  \Delta \mathcal{V}}   c'_{E_\mu,\alpha}
\end{equation}
  where we must divide by the 3D-3V phase-space volume $\Delta V  \Delta \mathcal{V}$.
  This procedure is repeated for each velocity bin  $\V{v}_{lmn}$ to construct the full alternative field-particle correlation at position $\V{r}_{ijk}$, given by 
  $C'_{\V{E}}(\V{r}_{ijk},\V{v})$, where the correlation is known at the discrete velocity bin centers $\V{v}_{lmn}$.

\item Finally, to obtain standard field-particle correlation  $C_{E_\mu}(\V{r}_{ijk},\V{v})$   at  position $\V{r}_{ijk}$ for electric field component $E_\mu$, we employ the velocity-space derivatives along velocity coordinate $v_\mu$ using \eqref{eq:cprimetoc}, 
  \begin{equation}
  C_{E_\mu}(\V{r}_{ijk},\V{v})=  -\frac{v_\mu}{2} \frac{\partial  C'_{E_\mu}(\V{r}_{ijk},\V{v})} {\partial v_\mu} + \frac{C'_{E_\mu}(\V{r}_{ijk},\V{v})} {2}
\end{equation}
 The velocity space derivative is computed using a centered, second-order finite difference, for example along the $v_x$ velocity dimension,
   \begin{equation}
     \frac{\partial  C'_{E_\mu}(\V{r}_{ijk},v_{x,l},v_{y,m},v_{z,n})} {\partial v_x}
     = \frac{ C'_{E_\mu}(\V{r}_{ijk},v_{x,l+1},v_{y,m},v_{z,n}) -
       C'_{E_\mu}(\V{r}_{ijk},v_{x,l-1},v_{y,m},v_{z,n})}{ v_{x,l+1}-v_{x,l-1}}.
\end{equation}
 For the velocity-space end points along each dimension at $\pm v_{max}$, a first-order, finite difference scheme is used to compute the velocity derivative.
   
\item   This procedure is repeated for each component $\mu=x,y,z$ of the electic field $\V{E}$ to obtain the full standard field-particle correlation  $C_{\V{E}}(\V{r}_{ijk},\V{v})$   at  position $\V{r}_{ijk}$
\begin{equation}
 C_{\V{E}}(\V{r}_{ijk},\V{v},t) = \sum_\mu  C_{E_\mu}(\V{r}_{ijk},\V{v})
\end{equation}

\item Note that when the correlation $ C_{\V{E}}(\V{r}_{ijk},\V{v},t)$ is integrated over all velocity space, the result is the integrated rate of change of spatial energy density  by the electric field over the spatial subvolume $\Delta V_{ijk}$,
\begin{equation}
 \int  d^3 \V{v} \, \, C_{\V{E}}(\V{r}_{ijk},\V{v},t) = \V{j}(\V{r}_{ijk},t) \cdot \V{E}(\V{r}_{ijk},t)
\end{equation}

\end{enumerate}

For the \texttt{dHybridR} simulation analyzed here, the weighting for
each PIC macroparticle is simply $W_\alpha=1$, so that the number of
macroparticles in a 3D-3V phase-space volume $\Delta V \Delta
\mathcal{V}$ is simply $N^{ijk}_{lmn}$.  Since the most important
spatial dimension in the study of particle energization at shocks is
the normal direction $x$ through the shock, we choose the spatial
subvolume to construct our velocity distribution to span a normal
range $\Delta x =d_i/4$ and the full transverse extent of the
simulation domain $\Delta y=L_y$ and $\Delta z=L_z$.

\section{Computing Shock Velocity}
\label{appendix:shocktrack}
\begin{figure} 
    \centering
    \includegraphics[width=1\textwidth]{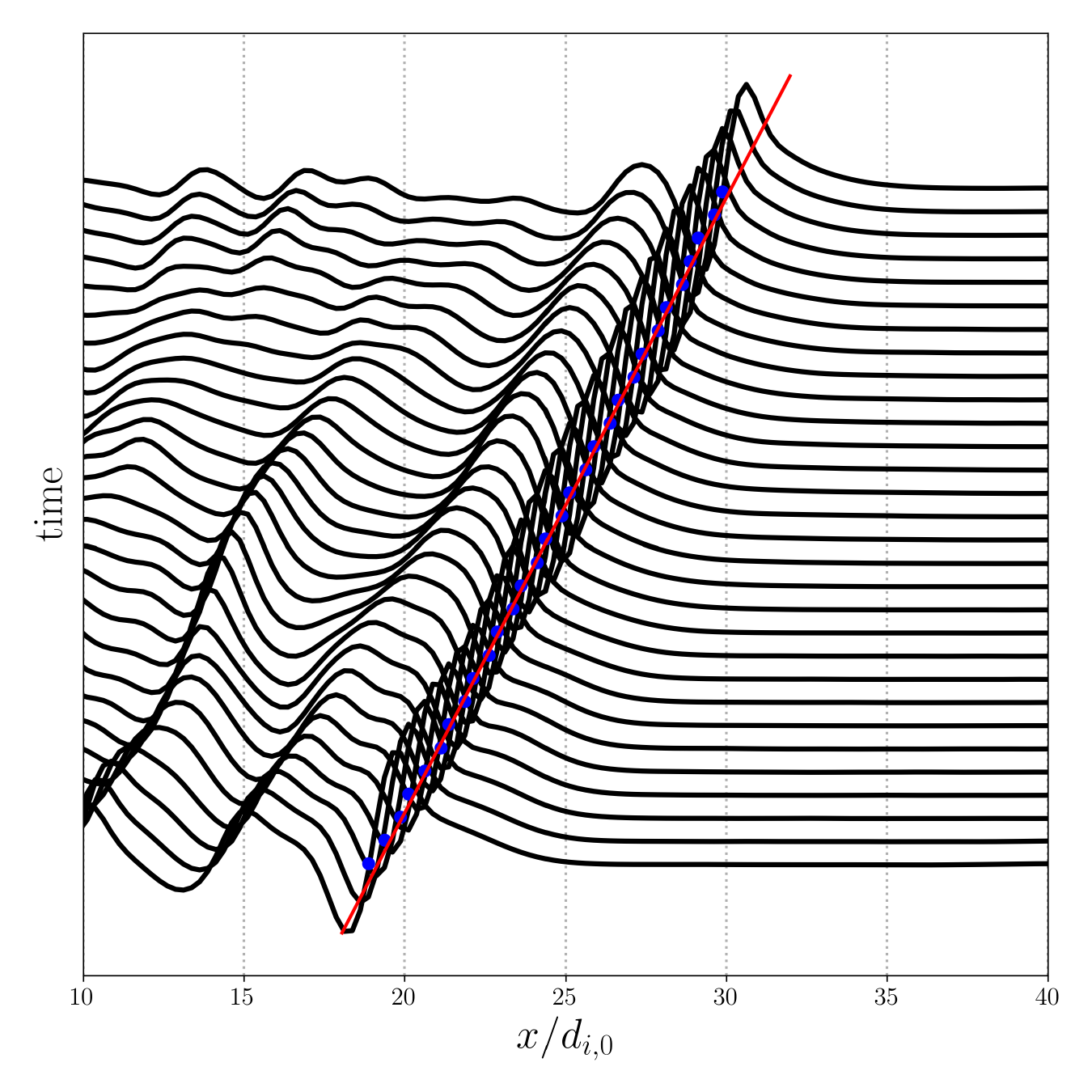}
    \caption{A timestack plot of the normal profiles of the average cross-shock electric field $\overline{E}_x(x)$ plotted every $\Delta t \, \Omega_{i,0} =0.2$, where the vertical displacement of each trace indicates the time evolution. The blue points indicate the shock position $x_s(t)$ at the zero crossing of $\overline{E}_x(x)$.  The red line shows the linear fit used to compute the shock velocity $U_s$, confirmation our procedure returns an approximately constant shock velocity in the simulation frame.}
    \label{fig:shocktrack}
\end{figure}

In the \texttt{dHybridR} \citep{haggerty2019dhybridr} shock simulation analyzed here, the upstream incoming flow in the $-\hat{x}$ direction impacts a reflecting wall at $x=0$, generating a shock that propagates in the $+\hat{x}$ direction in the simulation frame of reference, which is the frame of reference in which the average downstream bulk plasma flow is zero (the downstream rest frame).  We choose to apply the field-particle correlation technique in the shock-rest frame, so it is necessary to determine the the velocity of the shock front in the simulation frame.  For a moderately supercritical ($M_A \lesssim 12$) shock with $45^\circ \lesssim \theta_{Bn} \lesssim 90^\circ$, we find the hybrid simulations produce a predictable structure along the normal direction in the average cross-shock electric field, $\overline{E}_x(x)$.  As illustrated by the red-dashed line in the lower panel of \figref{fig:fieldscloseup}, $\overline{E}_x(x)$ rises from approximately zero far upstream to a local maximum within the ramp of the shock, before crossing through zero at approximately the same location along the normal as the peak of the magnetic field overshoot, seen in  $\overline{B}_z(x)$ (blue dotted) in the upper panel. We define the position $x_s$ of this zero crossing $\overline{E}_x(x_s) = 0$ to be the position of the shock for the purpose of computing the shock propagation velocity. As illustrated in \figref{fig:shocktrack}, we select a subset of snapshots in time after the shock has formed, computing a linear fit to the position $x_s(t)$ to determine the approximate velocity of the shock $U_s$ throughout the simulation. Our analysis returns a shock velocity in the $+\hat{x}$ direction of magnitude $U_s/v_A=1.88 \pm 0.01$, under the assumption of residual normality.  We then Lorentz transform the electromagnetic fields and the velocity distribution functions to a new frame of reference moving at $U_s$.  In the resulting \emph{shock-rest frame of reference}, the upstream plasma flow has an \Alfven Mach number $M_A=7.88$.

We compare our determination of the \Alfven Mach number in the shock-rest frame, $M_A=7.88$, to the value predicted by the MHD Rankine-Hugoniot jump conditions \citep{Burgess:2015}. The \Alfven Mach number in the shock-rest frame $M_A$ can be related to the upstream inflow velocity in the simulation (downstream rest) frame $U'_1/v_A=6$ by the relation $M_A = r(U'_1/v_A)/(r-1)$, where the MHD Rankine-Hugoniot jump conditions yield the shock compression ratio $r \equiv \rho_2/\rho_1 = U_1/U_2 =r(M_A,\theta_{Bn},\beta)$ in terms of the upstream parameters $M_A$, $\theta_{B_n}=45^\circ$, and $\beta= \beta_i+\beta_e = 2$.  We adjust $M_A$ in the shock-rest frame iteratively until we find agreement, yielding a compression ratio $r=3.62$ and a Mach number $M_A=8.29$.  Our measured value of $M_A=7.88$ is about 5\% lower than the MHD prediction, which is not unexpected for several reasons: (i) particles that reflect at the shock and escape upstream lower the effective incoming flow velocity; (ii) the MHD description does not capture the deviation between ion and electron dynamics in the collisionless system; and (iii) the collisionless dynamics can yield deviations of the effective adiabatic index from the strongly collisional MHD value for a fully ionized, hydrogenic plasma of $\gamma=5/3$ \citep{haggerty2020kinetic,caprioli2020kinetic}.

\end{document}